\documentclass[traditabstract]{aa} % for the abstract without structuration 

\usepackage{graphicx}
\usepackage{rotating}
\usepackage{txfonts}
\usepackage{natbib}
\bibpunct{(}{)}{;}{a}{}{,}

\newcommand{\um}{$\mu$m}

\newcommand{\kms}{km\thinspace s$^{-1}$}

\def\degr{\hbox{$^\circ$}}

\def\arcsec{\hbox{$^{\prime\prime}$}}
\def\utw{\smash{\rlap{\lower5pt\hbox{$\sim$}}}}
\def\udtw{\smash{\rlap{\lower6pt\hbox{$\approx$}}}}

\def\Msun{\hbox{\it M$_\odot$}}

\def\Mbol{\hbox{\it M$_{bol}$}}

\newcommand{\Ks}{{\it K$_{\rm s}$}}

\newcommand{\Aks}{{\it A$_{\it K_{\rm s}}$}}
\newcommand{\Ak}{{\it A$_{\it K}$}}

\def\simgr{\mathrel{\hbox{\rlap{\hbox{\lower4pt\hbox{$\sim$}}}\hbox{$>$}}}}
\def\Vlsr{\hbox{\it V$_{\rm lsr}$}}

\begin{document}

\title{86 GHz SiO maser survey of late-type stars in the Inner Galaxy. 
IV. SiO emission and infrared  data for sources in the 
Scutum and Sagittarius-Carina arms, $20^\circ<l<50^\circ$
\thanks{ 
The full Table 3 is only available in electronic form at
the CDS via anonymous ftp to
cdsarc.u-strasbg.fr (130.79.128.5)
or via http://cdsweb.u-strasbg.fr/cgi-bin/qcat?J/A+A/xxx/xxx.}
}
   \subtitle{}

\author{Messineo, M. 
\inst{1,2} \thanks{MM is currently employed by the University of Science and Technology of China. 
This works was partially carried out during her PhD thesis (2000-2004) in Leiden.}
\and
H.~J., Habing
\inst{2}
\and
L.~O., Sjouwerman 
\inst{3}
\and
A., Omont, 
\inst{4}
\and
K.~M., Menten
\inst{5}
}
          
\institute{
Key Laboratory for Researches in Galaxies and Cosmology, University of Science and Technology of China, 
Chinese Academy of Sciences, Hefei, Anhui, 230026, China
\email{messineo@ustc.edu.cn}
\and
Leiden Observatory, PO Box 9513, 2300 RA Leiden, The Netherlands
\and
National Radio Astronomy Observatory, PO Box 0, Socorro NM 87801, USA
\and          
Institut d'Astrophysique de Paris, CNRS, 98bis boulevard Arago, 75014 Paris, France
\and 
Max-Planck-Institut f\"ur Radioastronomie, Auf dem H\"ugel 69, D-53121 Bonn, Germany
}

\date{Received    2011/ Accepted }
\titlerunning{86 GHz SiO maser survey in the Inner Galaxy.IV  }
\authorrunning{Messineo et al.}

\abstract
{
We present an 86 GHz SiO ($v = 1, J = 2 \rightarrow 1$) maser search toward late-type 
stars located within $|b|<0.\!\!^{\circ}$5 and $20^\circ<l<50^\circ$.
This search is an extension at longer longitudes of a previously published work. 
We selected 135 stars from the MSX catalog using color and flux criteria and detected 92 
(86 new detections). The detection rate is 68\%, the same as in our previous study.

The last few decades have seen the publication of several catalogs of point sources detected 
in infrared surveys (MSX,  2MASS,  DENIS, ISOGAL, WISE, GLIMPSE,  AKARI, and  MIPSGAL). 
We searched each catalog for data on the 444 targets  of our earlier survey and for the 
135 in the survey reported here. We confirm that, as anticipated, most of our targets have colors typical of 
oxygen-rich  asymptotic giant branch (AGB) 
stars. Only one target star  may have already left the AGB.  
Ten stars have colors typical of  carbon-rich stars, meaning a contamination of 
our sample with carbon stars  $\protect\la $ 1.7 \%.
}


\maketitle

\section{\label{introduction}Introduction}

 Asymptotic giant branch (AGB)  stars are rare, but they are among the brightest stars at 
infrared wavelengths. They lose mass at rates from 10$^{-9}$ \Msun\ $yr^{-1}$ up to 10$^{-4}$ \Msun\ $yr^{-1}$, 
and are often surrounded by circumstellar  envelopes where maser emission from SiO, H$_2$O, 
and OH may arise.  Maser observations  provide information on  the angular distribution 
and accurate line-of-sight velocities of AGB stars. A number of maser surveys have been 
carried out to measure stellar line-of-sight velocities toward the inner Galaxy 
\citep[e.g.,][]{lindqvist92, blommaert94, sevenster01, sevenster97a, sevenster97b, sjouwerman98, izumiura99, deguchi00a, deguchi00b}.  
These surveys were mostly aimed at the detection of  OH/IR stars \footnote{AGB
stars with OH maser emission in the 1612 MHz line, mostly
undetected at visual wavelengths, but bright in the IR.}. 
About 800 OH/IR stars were detected with the first
largescale blind surveys at 1612 MHz by 
\citet[e.g.,][]{sevenster97a, sevenster97b, sevenster01}. 
Kinematic modelings of that data allowed for constraints on the Galactic bar’s properties 
and yielded quantitative parameters of the bar \citep[e.g.,][]{sevenster99, debattista02, habing06}. 

Stars of the AGB also emit SiO maser emission at 43 and 86 GHz and more stars show this maser  
than that of OH \citep{habing06}. SiO-maser surveys have been carried out around the years 2000-2002 
with the Nobeyama telescope \citep[][]{deguchi00}. These surveys mainly targeted IRAS 
point sources and suffered from significant confusion near the Galactic plane. 
Large infrared surveys of the Galactic plane with less confusion than IRAS were 
already available at the beginning of the year 2000, when we embarked on a search 
for 86 GHz SiO masers directed at targets selected from the ISOGAL 
\citep{omont03, schuller03} and MSX catalogs \citep{egan99,egan03}. 
We detected 255 SiO maser lines in 444 targets in the area 
$|b|<0.\!\!^{\circ}$5 and $-4^\circ<l<+30^\circ$ 
\citep[][]{messineo02}.\defcitealias{messineo03_2}{Paper\,II} 
The targets were selected to be complementary to the previous OH/IR  
surveys, in other words,  sources with the reddest mid- and near-infrared colors were excluded.\\

In 2003, an additional dataset of SiO masers at 86 GHz was obtained to extend the 
survey to longer longitudes ($20^\circ<l<50^\circ$).  A kinematic analysis of this dataset, 
together with that of \citet{messineo02}, is presented in \citet{habing06}. 
In this paper we publish the list of these 92  SiO maser detections between $20^\circ<l<50^\circ$. 
We also discuss the infrared properties of all our 444 previous targets and the 135 new 
targets as they are found in the new infrared surveys, MSX,  2MASS,  DENIS,  ISOGAL, WISE, 
GLIMPSE,  AKARI, and  MIPSGAL.

In Sect.\  \ref{observations}, we present the target selection and radio observations 
of the 135 new targets, and in Sect.\ \ref{infrared} the identifications of counterparts 
from  available large infrared surveys of the 444 + 135 targets in the Galactic plane.   
In Sect. \ref{analysismaser}, we briefly describe the detection rate and properties of 
the newly detected SiO maser lines, and in Sect. \ref{analysisinfrared}  we discuss the infrared 
colors of all 579 targeted stars. Finally, a summary is provided in Sect. \ref{summary}.

\section{86 GHz SiO masers in the 2003 IRAM campaign}
\label{observations}

\begin{figure}
\begin{center}
\resizebox{0.8\hsize}{!}{\includegraphics[angle=0]{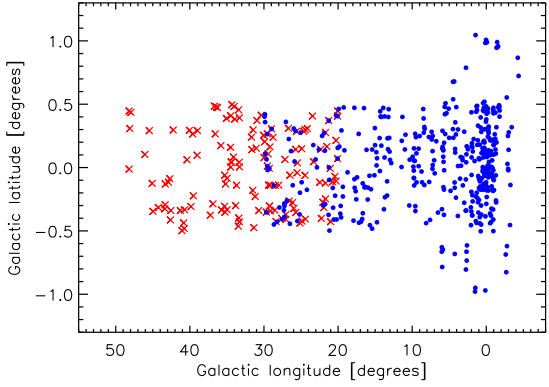}}
\caption{\label{fig.lb} Galactic coordinates of the new targets
(red crosses). 
For comparison the targets in \citet{messineo02}  (blue dots) are also shown. }
\end{center}
\end{figure}

In our previous paper, we  color-selected  bright  ($[15] < 3.4$ mag) stars in the ISOGAL [15] versus (\Ks$-[15]$) 
diagram, and  ISOGAL $[15]$ versus ($[7]-[15]$) diagram,  so as to exclude
the reddest sources since  those usually do not show SiO but OH maser emission; 
we defined a similar selection with MSX colors and fluxes.
Sources for the 2003 SiO observations were selected from Version 1.2 of the Midcourse Space 
Experiment point source catalog \citep[MSX-PSC,][]{egan99} 
by following the color criteria   of our previous paper. 
Flux densities in D-band range from 0.56 Jy (3.78 mag) to 2.12 Jy (2.34 mag). 
A number of 127 targets are observed for the first time;  eight targets 
are reobservations of targets in \citet{messineo02}.
Due to visibility limits from the IRAM 30-m telescope, there is  a lower limit to the 
target longitude of about $-4^\circ$.  
We restricted the latitude mostly to the limits $-$0\fdg5 and 0\fdg5. 
Locations of the targets are shown in Fig. \ref{fig.lb}.
Evidence  of variability is available from DENIS/2MASS near-infrared measurements; 
the majority of our targets are long-period variables \citep[see][]{messineo04}. 
  
The observations were carried out with the IRAM 30-m telescope on Pico Veleta in
Spain between October and November 2003, under the proposal number 021-03 
(30 hours).   Telescope settings and observational strategy were the same as described in 
\citet{messineo02}. Two receivers were used to observe the two orthogonal linear polarizations.  
To each receiver, we attached  the low-resolution analog filter bank with a resolution of 
3.5 \kms\ and a coverage of 890 \kms, and in parallel  the  autocorrelator with a 
resolution of  1.1 \kms\ and  a bandwidth 973 \kms. The IRAM beam at 86 GHz has a 
full width at half-maximum (FWHM) of 29\arcsec. The observations were made in wobbler 
switching mode, with the wobbler throw varying between 100\arcsec\ and 200\arcsec. 
Integration times ranged from 10 to 24 minutes per source; the average spectral noise 
is 0.012 K with a $\sigma=0.003$ K. The conversion factor from antenna temperatures to 
flux densities is 6.2 Jy K$^{-1}$. Typically, the SiO maser line has a width of a few \kms;  
therefore, the line  is not resolved in the spectra from the filter bank (one or two channels).  
We considered, as a detection, only lines simultaneously detected in the spectra from the 
filter bank and in those from the autocorrelator. 
Data analysis was carried out with the CLASS package within the GILDAS software.
Maser parameters, such as velocities, FWHMs, and integrated area below the line emission (A), 
were estimated by fitting the detected lines  with a Gaussian function.

Table \ref{table:detections} lists the 92 detections at 86 GHz  (86 new detections), 
while Table \ref{table:nondetections} lists the 43 non-detected targets 
(41 observed for the first time). The spectra of the detected targets are shown 
in Fig.\ \ref{spectra}. Since the lists by \citet{messineo02} contain 444 targets, 
Table \ref{table:detections} begins with the identifier number 445.

\begin{table*}\renewcommand{\arraystretch}{0.8}
\caption{\label{table:detections} Sources with detected 86 GHz SiO maser emission.
}
\medskip  
\ \\ 
$$
{\tiny
\begin{array}{llrrcccrll} 
\noalign{\smallskip}  
\noalign{\hrule} 
\noalign{\smallskip}  
 {\rm ID} & {\rm MSX ID }   &{\rm V}$$_{\rm LSR}$$ & {\rm T} $$_{\rm a}$$ & {\rm rms} & {\rm A}                   &    {\rm FWHM}         & {\rm Obs. Date} &{\rm SIMBAD~alias} \\ 
          &                 &[{\rm km\ s}$$^{-1}$$]& [{\rm K}]            &[{\rm K}]  & [{\rm K\ km\ s}$$^{-1}$$] &[{\rm km\ s}$$^{-1}$$] & [{\rm ddmmyy}]  & \\ 
\noalign{\smallskip}  
\noalign{\hrule} 
   445 &  G020.0327$+$00.4456 &     129.7&  0.058&  0.011&   0.29$$\pm$$   0.03 &    4.8 $$\pm$$   0.6 &          031003 &                                                  \\
   446 &  G020.2682$+$00.2141 &     102.5&  0.053&  0.014&   0.33$$\pm$$   0.05 &    6.2 $$\pm$$   1.0 &          031003 &                                                  \\
   447 &  G020.2609$-$00.1068 &     123.2&  0.080&  0.009&   0.45$$\pm$$   0.04 &    5.4 $$\pm$$   0.5 &          031003 &                                                  \\
   448 &  G021.0506$-$00.0063 &      28.7&  0.293&  0.010&   1.22$$\pm$$   0.03 &    4.4 $$\pm$$   0.1 &          031003 &                                                  \\
   449^{\rm a} &  G021.6986$+$00.2768 &      72.1&  0.139&  0.014&   1.11$$\pm$$   0.06 &    8.2 $$\pm$$   0.5 &          101003 &               {\rm 2MASSJ}18294455$$-$$0951203     \\
   450 &  G021.5015$+$00.1654 &      53.7&  0.047&  0.010&   0.14$$\pm$$   0.03 &    2.6 $$\pm$$   0.6 &          031003 &                                                  \\
   451 &  G021.0134$-$00.4268 &     155.8&  0.045&  0.010&   0.18$$\pm$$   0.03 &    3.8 $$\pm$$   0.7 &          031003 &                                                  \\
   452 &  G021.7102$-$00.1233 &      96.0&  0.062&  0.011&   0.84$$\pm$$   0.06 &   13.3 $$\pm$$   1.1 &          101003 &                                                  \\
   453 &  G021.6933$-$00.2472 &     135.1&  0.063&  0.013&   0.27$$\pm$$   0.04 &    4.1 $$\pm$$   0.7 &          101003 &                                                  \\
   454 &  G021.9071$-$00.3158 &     110.2&  0.173&  0.010&   0.57$$\pm$$   0.03 &    3.1 $$\pm$$   0.2 &          101003 &                                                  \\
   455 &  G023.4435$+$00.4059 &     112.3&  0.069&  0.010&   0.25$$\pm$$   0.04 &    3.6 $$\pm$$   0.7 &   111003/221003 &                                                  \\
   456 &  G022.4636$-$00.1162 &     111.2&  0.047&  0.009&   0.16$$\pm$$   0.03 &    3.5 $$\pm$$   0.8 &          101003 &                                                  \\
   457 &  G022.2467$-$00.4010 &      22.2&  0.045&  0.010&   0.30$$\pm$$   0.05 &    9.1 $$\pm$$   1.6 &          101003 &                                                  \\
   458 &  G023.3004$+$00.0466 &     128.6&  0.094&  0.008&   0.50$$\pm$$   0.04 &    6.4 $$\pm$$   0.7 &          101003 &                                                  \\
   459 &  G024.1345$+$00.3074 &     141.7&  0.113&  0.020&   0.92$$\pm$$   0.11 &   10.4 $$\pm$$   1.4 &          111003 &                                                  \\
   460 &  G024.4277$+$00.2976 &      52.6&  0.083&  0.009&   0.77$$\pm$$   0.05 &   13.2 $$\pm$$   0.9 &          101003 &                                                  \\
   461 &  G024.4455$-$00.0612 &      53.7&  0.104&  0.014&   0.47$$\pm$$   0.05 &    4.1 $$\pm$$   0.6 &          121003 &                                                  \\
   462^{\rm b}  &  G025.6246$+$00.2846 &      33.0&  0.081&  0.011&   0.57$$\pm$$   0.05 &    7.8 $$\pm$$   0.9 &          121003 &           ${\rm IRAS 18343$-$0624}$    \\
   463 &  G025.4894$+$00.0372 &     $-$30.0&  0.047&  0.010&   0.21$$\pm$$   0.03 &    4.0 $$\pm$$   0.5 &   121003/031103 &                                                  \\
   464 &  G025.7324$+$00.1434 &      $-$0.7&  0.064&  0.008&   0.18$$\pm$$   0.03 &    2.7 $$\pm$$   0.5 &          111003 &                                                  \\
   465 &  G025.5134$-$00.0413 &      55.8&  0.129&  0.012&   0.45$$\pm$$   0.04 &    3.6 $$\pm$$   0.4 &          121003 &                                                  \\
   466 &  G024.8618$-$00.4230 &       1.5&  0.034&  0.009&   0.18$$\pm$$   0.03 &    5.9 $$\pm$$   1.2 &          121003 &                                                  \\
   467 &  G026.6929$+$00.3095 &      17.8&  0.055&  0.012&   0.15$$\pm$$   0.03 &    2.4 $$\pm$$   0.8 &          121003 &                                                  \\
   468 &  G025.6894$-$00.3543 &     $-$82.1&  0.071&  0.008&   0.28$$\pm$$   0.03 &    3.9 $$\pm$$   0.5 &          101003 &                                                  \\
   469 &  G025.9663$-$00.3220 &     116.7&  0.038&  0.009&   0.40$$\pm$$   0.07 &   12.7 $$\pm$$   2.2 &          121003 &                                                  \\
   470 &  G026.4890$-$00.0598 &      75.4&  0.086&  0.012&   0.67$$\pm$$   0.06 &    8.0 $$\pm$$   0.9 &          121003 &                                                  \\
   471 &  G026.1404$-$00.2768 &      72.1&  0.074&  0.012&   0.20$$\pm$$   0.03 &    2.4 $$\pm$$   0.5 &          121003 &                                                  \\
   472^{\rm c} &  G027.1520$+$00.0642 &     $-$39.8&  0.145&  0.012&   0.90$$\pm$$   0.04 &    5.6 $$\pm$$   0.3 &          221003 &                               {\rm LGB99}$-$37     \\
   473 &  G026.5920$-$00.2645 &      97.1&  0.106&  0.011&   0.57$$\pm$$   0.05 &    6.0 $$\pm$$   0.7 &          121003 &                                                  \\
   474 &  G027.0475$-$00.4083 &      67.8&  0.060&  0.011&   0.50$$\pm$$   0.07 &   11.8 $$\pm$$   2.2 &   211003/231003 &                                                  \\
   475 &  G028.0060$-$00.1391 &      76.5&  0.056&  0.012&   0.30$$\pm$$   0.05 &    6.0 $$\pm$$   1.1 &   221003/031103 &                                                  \\
   476 &  G027.5895$-$00.4022 &      76.5&  0.085&  0.013&   0.52$$\pm$$   0.06 &    7.3 $$\pm$$   1.2 &          221003 &                                                  \\
   477 &  G029.1788$+$00.1627 &      40.6&  0.111&  0.013&   0.37$$\pm$$   0.04 &    3.5 $$\pm$$   0.5 &          221003 &                                                  \\
   478 &  G028.3534$-$00.4268 &      99.3&  0.152&  0.031&   0.70$$\pm$$   0.12 &    4.7 $$\pm$$   1.1 &          211003 &                                                  \\
   479 &  G029.4346$+$00.1071 &      60.2&  0.068&  0.012&   0.38$$\pm$$   0.05 &    6.2 $$\pm$$   1.0 &          231003 &                                                  \\
   480 &  G030.0440$+$00.2546 &     113.4&  0.187&  0.013&   0.42$$\pm$$   0.03 &    2.1 $$\pm$$   0.1 &          231003 &                                                  \\
   481^{\rm d} &  G029.1249$-$00.3432 &       9.1&  0.042&  0.010&   0.40$$\pm$$   0.06 &   11.7 $$\pm$$   2.0 &   221003/031103 &                                  {\rm F3S05}     \\
   482 &  G029.9674$+$00.0786 &     $-$46.3&  0.309&  0.013&   1.48$$\pm$$   0.05 &    5.7 $$\pm$$   0.3 &          231003 &                                                  \\
   483^{\rm e} &  G029.2599$-$00.3034 &      66.7&  0.269&  0.013&   1.03$$\pm$$   0.05 &    4.3 $$\pm$$   0.3 &          221003 &                                    {\rm S}23     \\
   484 &  G030.9389$+$00.3722 &      70.0&  0.066&  0.013&   0.46$$\pm$$   0.06 &    7.3 $$\pm$$   1.1 &          231003 &                                                  \\
   485 &  G031.4550$+$00.2974 &      67.8&  0.068&  0.012&   0.21$$\pm$$   0.03 &    3.0 $$\pm$$   0.6 &          231003 &                                                  \\
   486 &  G031.6752$+$00.2396 &      92.8&  0.093&  0.012&   0.85$$\pm$$   0.06 &   10.6 $$\pm$$   0.9 &          301003 &                                                  \\
   487 &  G030.7034$-$00.3387 &      65.6&  0.127&  0.013&   0.70$$\pm$$   0.06 &    5.9 $$\pm$$   0.6 &          231003 &                                                  \\
   488 &  G031.1661$-$00.2286 &      31.9&  0.100&  0.011&   0.25$$\pm$$   0.03 &    2.2 $$\pm$$   0.3 &          231003 &                                                  \\
   489 &  G031.4963$-$00.1798 &      99.3&  0.077&  0.012&   0.47$$\pm$$   0.06 &    9.8 $$\pm$$   1.2 &          231003 &                                                  \\
\noalign{\smallskip}  
\noalign{\hrule} 
\end{array} 
}
$$ 
\begin{list}{}{}
\item[{\bf Notes.}]
($a$)~Star \#449  corresponds to  2MASSJ18294455-0951203, which is classified 
as an  M9-10I/III long period variable \citep{halpern07}.~
($b$)~Stars \#243, \#462,  \#505, and \#511 are known SiO maser emitters \citep{messineo02,deguchi04,deguchi10}.
($c$)~Star \#472  coincides with  [LGB99]37, an M9I/III star \citep{lopez-corredoira99};
($d$)~star \#481 (SiO \Vlsr=9.1 \kms) corresponds to [NGM2011]F3S05, a giant M7III, detected by   
\citet[ \Vlsr=$-37$ \kms, ][]{negueruela11}.
($e$)~Star \#483  is a candidate RSG ([CND2009]S23) in the cluster RSGC3 \citep{clark09}.                   %;;;asap  repeated detection 66.75kms 
\end{list}
\end{table*}

\addtocounter{table}{-1}
\begin{table*}\renewcommand{\arraystretch}{0.8}
\caption{ Continuation of Table \ref{table:detections}.}
\medskip  
\ \\ 
$$
{\tiny
\begin{array}{llrrcccrll} 
\noalign{\smallskip}  
\noalign{\hrule} 
\noalign{\smallskip}  
 {\rm ID} & {\rm MSX ID }   &{\rm V}$$_{\rm LSR}$$ & {\rm T} $$_{\rm a}$$ & {\rm rms} & {\rm A}                   &    {\rm FWHM}         & {\rm Obs. Date} &{\rm SIMBAD~alias} \\ 
          &                 &[{\rm km\ s}$$^{-1}$$]& [{\rm K}]            &[{\rm K}]  & [{\rm K\ km\ s}$$^{-1}$$] &[{\rm km\ s}$$^{-1}$$] & [{\rm ddmmyy}]  & \\ 
\noalign{\smallskip}  
\noalign{\hrule} 
   490 &  G031.3555$-$00.4748 &      66.7&  0.033&  0.010&   0.36$$\pm$$   0.05 &   12.4 $$\pm$$   1.7 &   231003/031103 &                                                  \\
   491 &  G033.4182$+$00.4558 &      31.9&  0.156&  0.014&   1.03$$\pm$$   0.09 &   14.6 $$\pm$$   1.5 &          301003 &                                                  \\
   492 &  G033.3057$+$00.3923 &      78.6&  0.199&  0.014&   1.04$$\pm$$   0.05 &    5.1 $$\pm$$   0.3 &          301003 &                                                  \\
   493 &  G034.0030$+$00.4828 &      78.6&  0.125&  0.014&   0.99$$\pm$$   0.08 &   11.1 $$\pm$$   1.2 &          301003 &                                                  \\
   494 &  G033.3397$+$00.0420 &      80.8&  0.058&  0.010&   0.12$$\pm$$   0.02 &    2.0 $$\pm$$   0.4 &          301003 &                                                  \\
   495 &  G033.3037$+$00.0059 &      84.1&  0.131&  0.014&   0.51$$\pm$$   0.05 &    3.8 $$\pm$$   0.5 &          301003 &                                                  \\
   496 &  G034.3200$+$00.4951 &      90.6&  0.365&  0.014&   1.05$$\pm$$   0.04 &    2.5 $$\pm$$   0.1 &          301003 &                                                  \\
   497 &  G034.4797$+$00.4087 &     159.0&  0.100&  0.013&   0.77$$\pm$$   0.06 &    8.2 $$\pm$$   0.7 &          301003 &                                                  \\
   498 &  G034.1053$+$00.0826 &      66.7&  0.156&  0.014&   0.72$$\pm$$   0.05 &    4.8 $$\pm$$   0.4 &          301003 &                                                  \\
   499 &  G033.6946$-$00.1398 &      11.3&  0.086&  0.014&   0.47$$\pm$$   0.05 &    6.1 $$\pm$$   0.9 &          301003 &                                                  \\
   500 &  G033.2976$-$00.4282 &      40.6&  0.055&  0.012&   0.25$$\pm$$   0.04 &    5.4 $$\pm$$   1.2 &          301003 &                                                  \\
   501 &  G034.5112$+$00.1592 &      50.4&  0.110&  0.014&   0.42$$\pm$$   0.05 &    3.9 $$\pm$$   0.6 &          301003 &                                                  \\
   502 &  G035.0567$+$00.3877 &      47.1&  0.155&  0.012&   1.46$$\pm$$   0.06 &    9.1 $$\pm$$   0.4 &          021103 &                                                  \\
   503 &  G034.4224$+$00.0444 &      33.0&  0.173&  0.013&   0.68$$\pm$$   0.04 &    3.8 $$\pm$$   0.3 &          301003 &                                                  \\
   504 &  G033.4608$-$00.4622 &      45.0&  0.227&  0.014&   0.67$$\pm$$   0.04 &    2.7 $$\pm$$   0.2 &          301003 &                                                  \\
   505{\rm ^b} &  G033.8179$-$00.3008 &      72.1&  0.108&  0.014&   0.40$$\pm$$   0.04 &    3.4 $$\pm$$   0.4 &          301003 &                     $[{\rm DNZ2010] Mc13$-$6}$     \\
   506 &  G034.9723$-$00.0836 &      71.0&  0.092&  0.013&   0.33$$\pm$$   0.04 &    3.1 $$\pm$$   0.4 &          301003 &                                                  \\
   507 &  G036.2534$+$00.4734 &      64.5&  0.080&  0.012&   0.44$$\pm$$   0.05 &    6.5 $$\pm$$   0.9 &          021103 &                                                  \\
   508 &  G035.2797$-$00.0600 &      37.4&  0.033&  0.009&   0.14$$\pm$$   0.03 &    4.2 $$\pm$$   0.8 &          021103 &                                                  \\
   509 &  G036.4216$+$00.4787 &      $-$2.8&  0.115&  0.011&   0.31$$\pm$$   0.03 &    2.3 $$\pm$$   0.2 &          021103 &                                                  \\
   510 &  G034.9710$-$00.3329 &      53.7&  0.140&  0.014&   0.80$$\pm$$   0.05 &    6.2 $$\pm$$   0.4 &          301003 &                                                  \\
   511{\rm ^b} &  G035.2548$-$00.3518 &       7.0&  0.078&  0.012&   0.67$$\pm$$   0.06 &    9.2 $$\pm$$   1.0 &          021103 &           ${\rm IRAS 18544+0150}$     \\
   512 &  G036.8735$-$00.2866 &      34.1&  0.122&  0.011&   0.46$$\pm$$   0.03 &    3.4 $$\pm$$   0.3 &          021103 &                                                  \\
   513 &  G037.2240$-$00.3764 &      81.9&  0.182&  0.011&   0.73$$\pm$$   0.04 &    4.0 $$\pm$$   0.4 &          021103 &                                                  \\
   514 &  G039.5957$+$00.2654 &      27.6&  0.048&  0.010&   0.31$$\pm$$   0.04 &    6.2 $$\pm$$   1.2 &          021103 &                                                  \\
   515 &  G040.0488$+$00.2936 &      76.5&  0.070&  0.013&   0.41$$\pm$$   0.07 &    6.9 $$\pm$$   1.7 &          021103 &                                                  \\
   516 &  G039.5220$-$00.2231 &      71.0&  0.037&  0.009&   0.18$$\pm$$   0.03 &    4.4 $$\pm$$   0.9 &          021103 &                                                  \\
   517{\rm ^f} &  G039.8380$-$00.3076 &      28.7&  0.163&  0.013&   0.54$$\pm$$   0.06 &    3.8 $$\pm$$   0.6 &          021103 &                        {\rm IRAS} 19027+0555     \\
   518 &  G040.7140$-$00.3327 &      80.8&  0.050&  0.008&   0.19$$\pm$$   0.02 &    3.6 $$\pm$$   0.5 &          021103 &                                                  \\
   519 &  G042.2422$+$00.2969 &      55.8&  0.176&  0.010&   0.90$$\pm$$   0.03 &    4.8 $$\pm$$   0.2 &          021103 &                                                  \\
   520 &  G040.7482$-$00.4813 &      67.8&  0.046&  0.009&   0.30$$\pm$$   0.04 &    8.2 $$\pm$$   1.1 &          021103 &                                                  \\
   521 &  G041.2734$-$00.3388 &      14.6&  0.037&  0.009&   0.20$$\pm$$   0.04 &    6.1 $$\pm$$   1.6 &          021103 &                                                  \\
   522 &  G041.0632$-$00.4985 &      36.3&  0.075&  0.012&   0.25$$\pm$$   0.04 &    2.9 $$\pm$$   0.6 &          021103 &                                                  \\
   523 &  G042.5589$-$00.3387 &      51.5&  0.189&  0.011&   0.68$$\pm$$   0.04 &    3.6 $$\pm$$   0.3 &          021103 &                                                  \\
   524 &  G042.6182$-$00.4128 &      60.2&  0.049&  0.009&   0.17$$\pm$$   0.03 &    3.5 $$\pm$$   0.7 &          021103 &                                                  \\
   525{\rm ^f} &  G043.3182$-$00.2932 &      62.4&  0.424&  0.015&   2.23$$\pm$$   0.05 &    5.1 $$\pm$$   0.1 &          031103 &                        {\rm IRAS} 19091+0900     \\
   526 &  G044.2448$-$00.3153 &      35.2&  0.101&  0.015&   0.80$$\pm$$   0.07 &    9.7 $$\pm$$   0.8 &          031103 &                                                  \\
   527 &  G045.4641$+$00.2914 &      36.3&  0.195&  0.017&   0.85$$\pm$$   0.06 &    4.5 $$\pm$$   0.5 &          031103 &                                                  \\
   528 &  G044.9698$-$00.3452 &      33.0&  0.095&  0.014&   0.30$$\pm$$   0.05 &    3.3 $$\pm$$   0.7 &          031103 &                                                  \\
   529 &  G046.0606$+$00.1034 &     $-$11.5&  0.090&  0.012&   0.64$$\pm$$   0.05 &    6.6 $$\pm$$   0.6 &          031103 &                                                  \\
   530 &  G047.9533$+$00.4354 &      60.2&  0.072&  0.016&   0.45$$\pm$$   0.09 &   10.4 $$\pm$$   2.5 &          031103 &                                                  \\
\hline
   416 &  G020.0732$+$00.4054 &     131.9&  0.159&  0.009&   0.48$$\pm$$   0.03 &    2.8 $$\pm$$   0.2 &          031003 &                        $[{\rm MHS2002}] 416$     \\
   418 &  G020.0667$+$00.0706 &      79.7&  0.074&  0.008&   0.22$$\pm$$   0.02 &    2.7 $$\pm$$   0.3 &          031003 &                        $[{\rm MHS2002}] 418$     \\
   243{\rm ^b} &  G021.2558$+$00.3704 &     110.2&  0.074&  0.008&   0.31$$\pm$$   0.03 &    4.0 $$\pm$$   0.4 &   101003/221003 &                        $[{\rm MHS2002}] 243$     \\
   432 &  G028.9721$+$00.3057 &      98.2&  0.038&  0.014&   0.61$$\pm$$   0.11 &   19.9 $$\pm$$   5.2 &          231003 &                        $[{\rm MHS2002}] 432$     \\
   435 &  G029.5447$+$00.2644 &      45.0&  0.082&  0.013&   0.31$$\pm$$   0.04 &    3.5 $$\pm$$   0.4 &          231003 &                        $[{\rm MHS2002}] 435$     \\
   440 &  G029.2733$-$00.0067 &     $-$19.1&  0.171&  0.013&   0.70$$\pm$$   0.04 &    4.3 $$\pm$$   0.3 &          231003 &                        $[{\rm MHS2002}] 440$     \\

\noalign{\smallskip}  
\noalign{\hrule} 
\end{array} 
}
$$ 
\begin{list}{}{}
\item[{\bf Notes.}]
($f$)~For stars \#517 (IRAS 19027+0555) and \#525 (IRAS 19091+0900)
there are  IRAS low resolution spectra  \citep[LRS,][]{kwok97};
they are classified as class I (incomplete) and class P 
(red continuum or presence of PAHs) respectively.
\end{list}
\end{table*}

\begin{table}[h]\renewcommand{\arraystretch}{0.8}
\caption{\label{table:nondetections} Sources not detected at 86 GHz.} 
\medskip  
\ \\ 
$$ 
{\tiny
\begin{array}{lcccll} 
\noalign{\smallskip}  
\noalign{\hrule} 
\noalign{\smallskip}  
 {\rm ID} & {\rm MSX ID }  &  {\rm rms}  &  {\rm Obs.  date}   \\ 
          &                & [{\rm K}] & [{\rm ddmmyy}]       \\ 
\noalign{\smallskip}  
\noalign{\hrule} 

   531 &  G020.4122-00.0703 & 0.010 &         031003 &                                                \\
   532 &  G020.7490-00.1120 & 0.008 &         031003 &                                                \\
   533 &  G022.3430+00.1438 & 0.010 &         101003 &                                                \\
   534{^\mathrm{*}}{^\mathrm{a}} &  G022.0265-00.2741 & 0.011 &         101003 &                \\
   535{^\mathrm{*}}{^\mathrm{b}} &  G024.6265+00.2856 & 0.009 &         121003 &                   \\
   536 &  G024.8993+00.1493 & 0.009 &         121003 &                                                \\
   537 &  G024.4084-00.4077 & 0.013 &         121003 &                                                \\
   538 &  G025.1373-00.4375 & 0.012 &         121003 &                                                \\
   539 &  G027.0540-00.3708 & 0.011 &  211003/231003 &                                                \\
   540 &  G028.7480+00.2394 & 0.013 &         221003 &                                                \\
   541 &  G028.4568-00.1409 & 0.013 &  211003/031103 &                                                \\
   542 &  G028.5241-00.3361 & 0.013 &         221003 &                                                \\
   543 &  G029.1718-00.1392 & 0.011 &         221003 &                                                \\
   544 &  G029.9008+00.2034 & 0.014 &         231003 &                                                \\
   545 &  G030.3701+00.4129 & 0.013 &         231003 &                                                \\
   546 &  G029.3767-00.1393 & 0.013 &         231003 &                                                \\
   547 &  G029.2366-00.3039 & 0.013 &         221003 &                                                \\
   548 &  G031.0465+00.1522 & 0.012 &         231003 &                                                \\
   549 &  G031.6255+00.1417 & 0.012 &         231003 &                                                \\
   550 &  G033.9410+00.3822 & 0.013 &         301003 &                                                \\
   551 &  G033.3780-00.3974 & 0.014 &         301003 &                                                \\
   552 &  G034.3524+00.0414 & 0.014 &         301003 &                                                \\
   553 &  G034.4886+00.0030 & 0.014 &         301003 &                                                \\
   554{^\mathrm{*}}{^\mathrm{d}} &  G035.1766-00.1296 & 0.012 &         021103 &   \\
   555 &  G035.8117+00.1731 & 0.011 &         021103 &                                                \\
   556 &  G036.6828+00.4850 & 0.009 &         021103 &                                                \\
   557 &  G035.3035-00.3060 & 0.012 &         021103 &                                                \\
   558 &  G036.5733+00.2658 & 0.011 &         021103 &                                                \\
   559 &  G035.7171-00.3353 & 0.009 &         021103 &                                                \\
   560 &  G038.9702+00.2921 & 0.012 &         021103 &                                                \\
   561 &  G039.0025+00.0971 & 0.012 &         021103 &                                                \\
   562 &  G040.3720+00.0615 & 0.011 &         021103 &                                                \\
   563 &  G040.8270-00.4366 & 0.011 &         021103 &                                                \\
   564 &  G042.6920-00.0883 & 0.015 &         031103 &                                                \\
   565 &  G042.8776-00.1158 & 0.015 &         031103 &                                                \\
   566 &  G043.3079-00.1290 & 0.015 &         031103 &                                                \\
   567 &  G043.3686-00.3254 & 0.015 &         031103 &                                                \\
   568{^\mathrm{*}}{^\mathrm{a}} &  G045.0643-00.1249 & 0.014 &         031103 &             \\
   569 &  G048.1911+00.4459 & 0.017 &         031103 &                                                \\
   570 &  G048.1018+00.3077 & 0.017 &         031103 &                                                \\
   571 &  G048.1924-00.0121 & 0.017 &         031103 &                                                \\
\hline
   441{^\mathrm{*}} &  G029.5298+00.0110 & 0.013 &         231003 &                         \\
   444{^\mathrm{*}}{^\mathrm{c}} &  G029.6750-00.3521 & 0.013 &         231003 &     \\
\noalign{\smallskip}  
\noalign{\hrule} 
\end{array} 
}
$$ 
\begin{list}{}{}
\item[{\bf Notes.}] (${\mathrm{*}}$)~Alias: 534 = ${\rm ISOGAL-PJ}183220.6-094910;$  $535= [{\rm NGM2010]~S13}$;
 441=$[{\rm MHS2002}] 441;$  444=[{\rm MHS2002}] 444/$[{\rm MZM2016}]68; $ 554={\rm IRAS} 18534+0152;
 568=${\rm ISOGAL-PJ}191416.8-104318.$~
($a$)~Stars \#534 and \#568 (non-detections) are reported as candidate YSOs by \citet{felli02}.
($b$)~Star \#535  is a candidate RSG ([NGM2010]S13) in the  
cluster Alicante 8 \citep{negueruela10}. %;;;;,asap  nondet   
($c$)~Star \#444  corresponds to MZM2016-68, an M2I \citep[][]{messineo16}.
($d$)~For star \#554 (IRAS 18534+0152) 
there is an  IRAS low resolution spectrum  \citep[LRS, featureless (F),][]{kwok97}.
\end{list} 
\end{table}

\section{Infrared counterparts to the targets searched for SiO masers}
\label{infrared}

We searched for available mid-infrared data for the 127 new targets 
as well as for the 444 targets of \citet{messineo02}. We used 
the final MSX release \citep[version 2.3, catalog V/114 in CDS,][]{egan03},
the ISOGAL catalog \citep{schuller03,omont03}, 
the GLIMPSE catalog \citep[catalog II/293 in CDS,][]{churchwell09,benjamin03},
the  AllWISE Data Release \citep[catalog II/328 in CDS,][]{wright10,cutri13},
the AKARI/IRC All-Sky Survey point source catalog v.1.0 \citep{ishihara10},
and the MIPSGAL 24 \um\ catalog by \citet{gutermuth15}.

The  targets had been selected from MSX catalog Version 1.2, which has an astrometric 
accuracy of $\sim 2$\arcsec\ \citep[resolution of 18\farcs3,][]{egan99}. 
For comparison, the targets selected from the ISOGAL catalog of \citet{messineo02}  
had a subarcsec accuracy
\citep[resolution of 1\arcsec,][]{messineo04}.
Since 2MASS sources have been astrometrically cross-correlated with WISE and GLIMPSE 
targets, we can now check the earlier 2MASS matches to the MSX sources (see Appendix \ref{Kmatches}). 
The MSX camera had a spatial resolution of 18\farcs3, while the beam of the IRAM  30-m 
telescope had a  FWHM of 29\arcsec. MSX counterparts  were found  within 14\farcs5 for 
566 of the 571 targets. Excluding the upper limits we have  280 targets with good 
measurements in all four MSX bands. MSX magnitudes are obtained adopting the 
following zero-points: 58.49 Jy in  the $A$ band (8.26$\mu$m), 26.51 Jy in 
$C$-band (12.12 $\mu$m), 18.29 Jy in $D$ band, (14.65 $\mu$m), and 8.80 Jy in 
$E$ band (21.41 $\mu$m) \citep{egan99}.
WISE has a resolution of 6\arcsec, and a final astrometric accuracy  
better than 0\farcs5. It imaged the sky at 3.4 \um\  ($W1$-band), 4.6 \um\  ($W2$-band), 
12 \um\  ($W3$-band), and 22 \um\ ($W4$-band) with a sensitivity of 0.08, 0.11, 1, 
and 6 mJy, and typical saturation thresholds at  0.18, 0.36, 0.88, and 12 Jy. 
93\% of our targets have unique WISE matches within 10\farcs0. A few MSX data points were 
resolved by WISE into several components; two WISE matches were available for 5\% of the 
targets. The closest match  was retained, which is also, in all but two cases, the brightest 
match in the $W3$ band (see Appendix \ref{Kmatches}).
For the Galactic Infrared Midplane Survey Extraordinaire (GLIMPSE), the Infrared Array 
Camera (IRAC) on board of the Spitzer Space Observatory was used to image the plane of 
the Galaxy at 3.6, 4.5, 5.8, and 8.0 \um\ with a spatial resolution of 1\farcs2, and a 
sensitivity of 0.2, 0.2, 0.6, and 0.4 mJy and saturation thresholds of 180, 190, 570, 
and 470 mJy. MSX measurements in the  8 \um\ (band-A) range  from 6.35 to 2.08 mag, 
with a Gaussian peak at 3.55 mag and a $\sigma=0.80$ mag. We considered only GLIMPSE 
stars  with  $[8.0] <  6.0$ mag as safe identifications. A second iteration with a 
fainter threshold was not needed. GLIMPSE counterparts  were found  for 76\% of the 
targets using a search radius of 10\arcsec\ and selecting the closest. 
For  63\% of the targets, the  MSX, WISE, and GLIMPSE selected counterparts are unique 
within the search radius.
For targets selected from the MSX catalog, with the 2MASS  magnitudes and positions, 
we searched for additional $IJK_s$ magnitudes 
from the DENIS  catalog \citep[using a search radius of 2\arcsec,][]{epchtein94}, and  
$IJK_s$, $7$ \um, and $15$ \um\ magnitudes from the combined DENIS-ISOGAL 
catalog \citep{omont03, schuller03}.
The 24 \um\ Spitzer  MIPSGAL survey has a spatial resolution of 6\arcsec\  
and a sensitivity from 1.3 mJy to 2 Jy; we found 518 counterparts within a search 
radius of 10\arcsec.
Finally, we searched for AKARI/IRC data at 9 \um\ and 18 \um\ \citep{ishihara10},
and found matches for 473 targets with a search radius of 3\farcs5.
Mid-infrared associations were verified with GLIMPSE charts %;shown in Fig.\ \ref{chartes} 
and  with a visual inspection of the stellar energy  distribution (SED). In addition, 
we  inspected UKIDSS images \citep{lucas08}, that were available for 536 of the 571 targets.
All but eight targets appeared as
single saturated stars in $K$-band and well centered on the 
2MASS position. 
The collected available infrared measurements of the targets are provided in 
Table \ref{table.magnitudes} together with the  positions from the 2MASS 
catalog \citep[][]{twomass}.

\section{SiO maser results}
\label{analysismaser}
We searched for SiO maser emission at 86 GHz toward our sample of 135 stars and detected 
92 SiO maser lines, of which 91 are new detections at 86 GHz. The resulting detection 
rate is 68\% and that compares very well with that in our earlier survey 
\citet[66\%,][]{messineo02}.

\subsection{ Single epoch detection rate and repeated observations  }
Star \#243 (\Vlsr=110.2 \kms) coincides with the entry \#243  in \citet[][]{messineo02}, 
and stars \#462 (\Vlsr=33.0 \kms),  \#505 (\Vlsr=72.1 \kms), and \#511 (\Vlsr=6.9 \kms)
were  detected at 43 GHz by \citet[][]{deguchi04} and \citet[][]{deguchi10}; 
for each star, measured velocities agree within 1 \kms. Interestingly, 
stars \#416, \#418, \#432, \#435, \#440, \#441, and \#444 were unsuccessfully 
searched for SiO emission by \citet{messineo02}. Five out of these seven stars not detected 
in the first epoch had a detectable flux during the second epoch of 86 GHz observations. 
This implies that  90\% of the targets in \citet{messineo02} are  probably masing stars.
In the second epoch the rms noise was lower by a factor of $\approx$1.3. Flux densities of 
SiO maser lines are known to vary in phase with the infrared amplitude of the stellar 
pulsation  \citep[e.g.,][]{alcolea99}. 
Since the selected sample is made out of long-period variables  \citep[][]{messineo04}, 
this single epoch detection rate provides only a lower limit to the actual number of 
stars capable of hosting SiO masers.

\subsection{FWHM of maser lines and candidate RSGs}
\label{fwhm}

\begin{figure}[h]
\begin{center}
\resizebox{1\hsize}{!}{\includegraphics[angle=0]{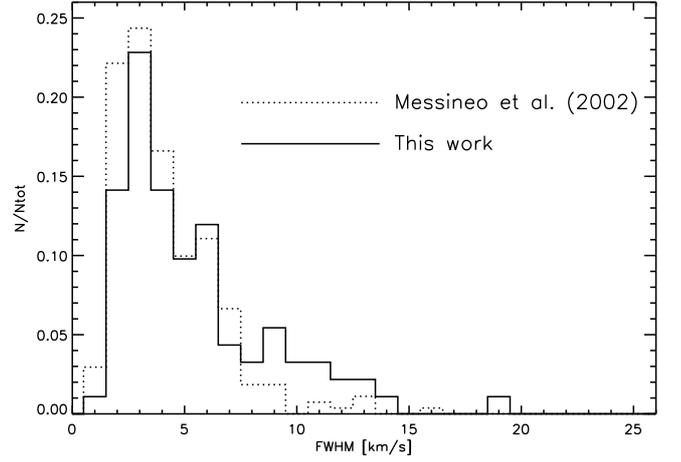}}
\end{center}
\caption{\label{fig:fwhm}   Histogram of the SiO maser line widths. 
The solid line shows the distribution of the SiO masers presented in this work; 
the dotted line those previously published by \citet{messineo02}. The numbers have been scaled to the total number (444) in the previous study and to the number 135 in the present study 
}
\end{figure}

The 86 GHz maser emission originates in the envelopes of O-rich (Mira-type) AGB stars, as well as in the envelopes 
of Red Supergiants (RSGs). Weak SiO maser emission is found in semiregular AGB stars 
\citep{alcolea90}.  The central star has an extended shell, generated by the strong 
pulsations where the SiO maser activity takes place. 
 Figure  \ref{fig:fwhm} shows the distribution of the FWHMs in comparison to our previous study. 
 Both distributions are 
compatible with each other, with the FWHM peaking at  4.9 \kms\  with a scatter of $\sigma=2.7$ \kms, 
 and the mean equivalent width (i.e., fitted area / peak)  peaking at 4.7 \kms\ with a scatter of 
$\sigma=2.0$ \kms. There is also a tail with broader FWHMs ( $>$ 7 \kms).
  
Cooler RSGs with large amplitudes are also strong SiO emitters, and their SiO maser 
intensity is comparable to that of long period variables \citep{alcolea90}. 
SiO masers associated with RSGs have larger widths \citep[e.g.,][]{verheyen12,lebertre90,alcolea90}. 
\citet{alcolea99} performed a six-year period monitoring of SiO maser lines and concluded that 
SiO maser emission contains several 1-2 \kms\ peaks over a range of 10 \kms\ 
in LPV-giants or 20-40 \kms\ in supergiants.
With this  statistical argument it should be possible to identify candidate RSGs.
However, in many cases lines were detected at the 2.5-3 $\sigma$ level and it is likely 
that most of the broad lines are the results of low signal-to-noise observations.
The quoted errors in Table \ref{table:detections} are the errors of the analytic 
fit with a Gaussian function. As expected, the broader lines are predominantly detected 
at lower antenna  temperatures.
By comparing the new sample with the old \citep{messineo02}, there is an excess of 
broader lines between 10$^\circ$ and 50$^\circ$ of longitudes. 14\% of masers 
located at longitudes $> 20^\circ$ has broader lines, but only 5\% with longitudes 
$< 20^\circ$. Between longitudes 25$^\circ$ and 35$^\circ$, at the near-end side of the bar, 
there is a concentration of  massive starburst clusters and those are rich in RSG stars 
\citep[e.g.,][and references therein]{messineo16}. A small excess contamination of RSGs 
is therefore expected at these longitudes (see Appendix \ref{mbol}).
The nature of objects with apparently broader lines is discussed  in Appendix B.

\section{Infrared color diagnostics}
\label{analysisinfrared}

\begin{figure}[h]
\begin{center}
\resizebox{0.8\hsize}{!}{\includegraphics[angle=0]{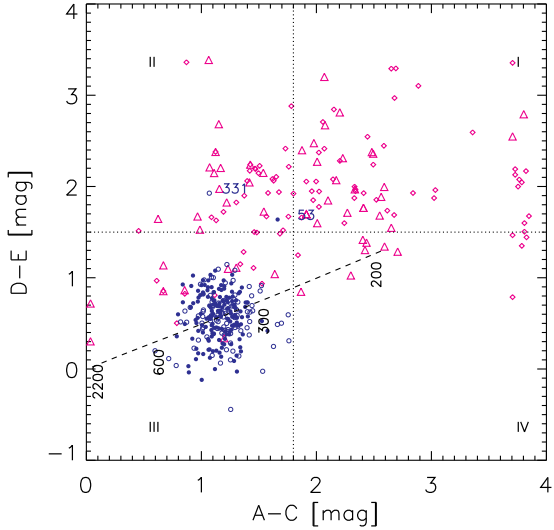}}
\end{center}
\caption{\label{fig:ppn} 
Diagram of targeted stars in the MSX $A-C$ versus $D-E$ colors. Upper limits and blends 
have been removed.  Targets are marked with blue filled circles (detections)  and  empty circles (non-detections).
As illustration we plot also the positions of astrophysically related objects. 
Triangles indicate the location of the PNe and diamond symbols that of post-AGB  
(transitional) objects provided by \citet{ortiz05}.
The long-dashed line marks the locus of a blackbody ($F_\nu$) with a 
temperature from 2200 K to 200 K.
}

\end{figure}

Because of their low latitudes and low fluxes, the targets have  upper limits at 60 \um\
and the classical IRAS color-color diagram cannot be used to characterise their envelopes 
\citep[Sect. 6 of ][]{messineo04}.
 \citet{sevenster02} derived useful diagnostics for OH/IR stars by analysing the MSX 
colors $[A]-[C]$  and $[D]-[E]$. In the plane defined  $D-E$ versus $A-C$ 
\citep[Fig. 3 of][]{sevenster02},   objects of different evolutionary status reside in different areas in 
the diagram. OH/IR stars reside in the area $A-C < 1.8$ mag and $D-E < 1.5$ mag 
(Quadrant III); planetary nebulae  (PNe) and post-AGB stars (objects in transition to  PNe)
populate the area $D-E > 1.5$ mag (Quadrants I and II). 
Our old and new targets are plotted in Fig. \ref{fig:ppn} in such a diagram, together with known post-AGB stars 
and PNe from \citet{ortiz05}. As expected, almost all of our objects are located in Quadrant III, which is the classical region 
of AGB stars. Only two targets are located in Quadrant II and could have left the AGB already: the SiO emitter \#53 and the non-detection \#331.  

\begin{figure}[h]
\begin{center}
\resizebox{1.0\hsize}{!}{\includegraphics[angle=0]{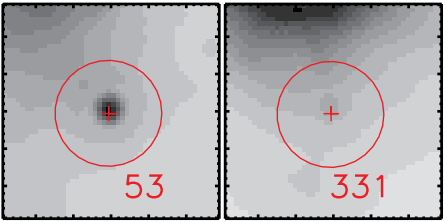}}
\end{center}
\caption{\label{mapmipsgal:ppn} MIPSGAL 24 \um\  $1^\prime \times 1^\prime$ maps 
of the two targets located in Quadrant II of Fig.\ \ref{fig:ppn}. Images have been 
bytescaled with a minimum at zero and a maximum at 500 MJy/sr  (\#53) or 
1900 MJy/sr  ( \#331). North is up and east to the left. The IRAM pointing 
positions are marked with circles of radius 14\farcs5. 
The final retained 2MASS positions  are marked with crosses. }
\end{figure}

We carefully inspected the quality of measurements and visually inspected the mid-infrared 
images of these candidate post-AGBs  (see Fig.  \ref{mapmipsgal:ppn}). 
Despite the good flags in the MSX bands both appear 
located in the halo  of a nearby bright 20 \um\ 
source, indicating a questionable photometric measurement.   
We conclude therefore that a contamination of our sample with post-AGB stars and PNe is negligible.


\begin{figure}[h]
\begin{center}
\resizebox{0.7\hsize}{!}{\includegraphics[angle=0]{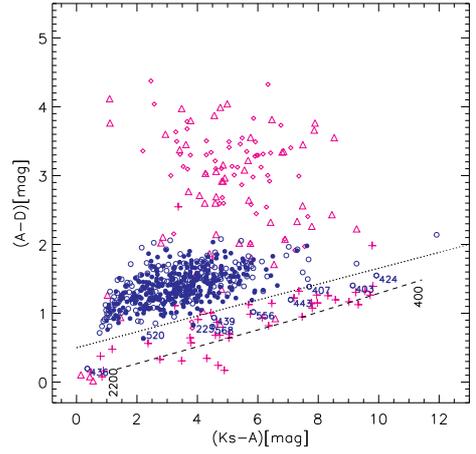}}
\end{center}
\caption{\label{fig:carbon} 2MASS-MSX \Ks$-A$ versus $A-D$ diagram of  targets. 
SiO maser detections are marked with blue filled circles and non-detections 
with blue empty circles.  Upper limits and blends have been removed. 
Identification numbers are as in Table \ref{table.magnitudes}.  The dotted line marks 
the separation between $O$-rich and $C$-rich shells found by \citet{ortiz05}. 
The long-dashed line represents a blackbody ($F_\nu$) with temperature from 2200 to 400 K. 
New candidate $C$-type stars from our survey (see text) are labeled.
For comparison \Ks$-A$ versus $A-D$ values of the carbon stars analyzed by \citet{ortiz05} 
are overplotted with magenta crosses, transitional objects with diamonds, and PNe with triangles. } 
\end{figure}

Color-color diagrams can be used also to search for different chemistry in our sample. \citet{ortiz05} showed that carbon stars can be 
identified in a 2MASS-MSX $A-D$ versus \Ks$-A$ diagram. In Fig.\ \ref{fig:carbon}, we show such a diagram with our targets from 
Table \ref{table.magnitudes} together with carbon stars, post-AGB stars, and PNe from \citet{ortiz05}. There are ten 
candidate $C$-rich stars in our sample, \#223, \#403,  \#407, \#424,  \#436, \#439,  
 \#443,   \#520, \#556, and \#568, which fall below the separation line dividing $O$-rich and $C$-rich stars in Fig. \ref{fig:carbon}. 
Two stars, \#223 and \#520, are masing at 86 GHz indicating an $O$-rich chemistry. Indeed, \#520 is also observed by AKARI and has 
a $[9]-[18]$ color typical of $O$-rich stars \citep{ishihara11}. We conclude that the non-simultaneity of the near-infrared and mid-infrared 
measurements yields some spurious $C$-rich stars.

\section{Summary}
\label{summary}

We have reported on 92 SiO maser detections at 86 GHz, which were made
with the IRAM 30-m single dish in 2003. 
This dataset along with the 271 detections by \citet{messineo02}
 was used as  probe of the Galactic barred potential in the work by \citet{habing06}.
The 2003 detections are located between 20$^\circ$ and 50$^\circ$ of longitude,
and yield a detection rate of 68\%, similar to that reported by \citet[][]{messineo02}.
Since no archival copy exists of this dataset, here, we make them available.

We performed identifications of the  135  targets   presented in this work
and  of the 444 of \citet{messineo02}
in the  MSX, WISE, MIPSGAL, GLIMPSE, 2MASS, DENIS, ISOGAL and AKARI catalogs.
The sample is made of Mira-like stars,
and a fraction of 74\% of stars shows photometric variability in at least one of the  band.

We used the collected measurements to confirm their $O$-rich nature, as  initially  
designed. For example, we analyze the $A-D$ versus \Ks$-A$ diagram 
to separate $O$- and $C$-rich stars, and found 1.7\%  of them could be C-rich stars.
The $D-E$ versus $A-C$ plane allows us to separate normal AGB stars from post-AGB stars and  PNe.
Only one masing star, \#53, is marginally deviating from the locus of normal AGB  stars in this diagram.

\begin{acknowledgements}
IRAM  is  supported   by  INSU/CNRS  (France),  MPG  (Germany) and   IGN   (Spain). 
ISOGAL is based on observations with ISO, an ESA project with instruments 
funded by ESA Member States (especially the PI countries: France, Germany, 
the Netherlands and the United Kingdom) and with the participation of ISAS and NASA.
DENIS is  a joint effort of several Institutes mostly located in Europe. It has
    been supported mainly by the French Institut National des
    Sciences de l'Univers, CNRS, and French Education Ministry, the
    European Southern Observatory, the State of Baden-Wuerttemberg, and
    the European Commission under networks of the SCIENCE and Human
    Capital and Mobility programs, the Landessternwarte, Heidelberg and
    Institut d'Astrophysique de Paris.
This publication makes use of data products from the Two Micron All Sky Survey, which is a joint project 
of the University of Massachusetts and the Infrared Processing and Analysis Center/California Institute 
of Technology, funded by the National Aeronautics and Space Administration and the National Science Foundation.
This work is based  on observations made with the Spitzer Space Telescope, 
which is operated by the Jet Propulsion Laboratory, California Institute of Technology under a contract with NASA.
This research made use of data products from the
Midcourse Space Experiment, the processing of which was funded by the Ballistic Missile Defence Organization with additional
support from the NASA Office of Space Science. This publication makes use of data products from
WISE, which is a joint project of the University of California, Los
Angeles, and the Jet Propulsion Laboratory/California Insti-
tute of Technology, funded by the National Aeronautics and
Space Administration.
This work is based in part on data obtained as part of the UKIRT Infrared Deep Sky Survey. 
This research has made use of the SIMBAD data base, 
operated at CDS, Strasbourg, France. This research made use of Montage, funded by the National Aeronautics and Space Administration's 
Earth Science Technology Office, Computational Technnologies Project, under Cooperative 
Agreement Number NCC5-626 between NASA and the California Institute of Technology. 
The code is maintained by the NASA/IPAC Infrared Science Archive.
AKARI is a JAXA project with the participation of the European Space Agency (ESA).
The work of MM from 2000 to 2004 was funded by the Netherlands
Research School for Astronomy (NOVA) through a network 2 Ph.D.
stipend. We thank  Dr. Sevenster M. N. for stimulating discussion on Galaxy morphology
and masing stars; she was a visiting fellow of the Leiden Observatory from 2002 to 2004.
This work was partially supported by  the  
Fundamental Research Funds for the Central Universities in China, and USTC grant KY2030000054.
The National Radio Astronomy Observatory is a facility of the National Science 
Foundation operated under cooperative agreement by Associated 
Universities, Inc. We thank the referees of this paper.
\end{acknowledgements}

%\bibliographystyle{aa}
%\bibliography{noteScutum.biblio}

\begin{thebibliography}{53}
\expandafter\ifx\csname natexlab\endcsname\relax\def\natexlab#1{#1}\fi

\bibitem[{{Alcolea} {et~al.}(1990){Alcolea}, {Bujarrabal}, \&
  {Gomez-Gonzalez}}]{alcolea90}
{Alcolea}, J., {Bujarrabal}, V., \& {Gomez-Gonzalez}, J. 1990, \aap, 231, 431

\bibitem[{{Alcolea} {et~al.}(1999){Alcolea}, {Pardo}, {Bujarrabal},
  {Bachiller}, {Barcia}, {Colomer}, {Gallego}, {G{\'o}mez-Gonz{\'a}lez}, {del
  Pino Cisneros}, {Planesas}, {del R{\'{\i}}o}, {Rodr{\'{\i}}guez-Franco}, {del
  Romero}, {Tafalla}, \& {de Vicente}}]{alcolea99}
{Alcolea}, J., {Pardo}, J.~R., {Bujarrabal}, V., {et~al.} 1999, \aaps, 139, 461

\bibitem[{{Benjamin} {et~al.}(2003){Benjamin}, {Churchwell}, {Babler}, {Bania},
  {Clemens}, {Cohen}, {Dickey}, {Indebetouw}, {Jackson}, {Kobulnicky},
  {Lazarian}, {Marston}, {Mathis}, {Meade}, {Seager}, {Stolovy}, {Watson},
  {Whitney}, {Wolff}, \& {Wolfire}}]{benjamin03}
{Benjamin}, R.~A., {Churchwell}, E., {Babler}, B.~L., {et~al.} 2003, \pasp,
  115, 953

\bibitem[{{Blommaert} {et~al.}(1994){Blommaert}, {van Langevelde}, \&
  {Michiels}}]{blommaert94}
{Blommaert}, J.~A.~D.~L., {van Langevelde}, H.~J., \& {Michiels}, W.~F.~P.
  1994, \aap, 287, 479

\bibitem[{{Cabrera-Lavers} {et~al.}(2006){Cabrera-Lavers}, {Garz{\'o}n},
  {Hammersley}, {Vicente}, \& {Gonz{\'a}lez-Fern{\'a}ndez}}]{cabrera06}
{Cabrera-Lavers}, A., {Garz{\'o}n}, F., {Hammersley}, P.~L., {Vicente}, B., \&
  {Gonz{\'a}lez-Fern{\'a}ndez}, C. 2006, \aap, 453, 371

\bibitem[{{Churchwell} {et~al.}(2009){Churchwell}, {Babler}, {Meade},
  {Whitney}, {Benjamin}, {Indebetouw}, {Cyganowski}, {Robitaille}, {Povich},
  {Watson}, \& {Bracker}}]{churchwell09}
{Churchwell}, E., {Babler}, B.~L., {Meade}, M.~R., {et~al.} 2009, \pasp, 121,
  213

\bibitem[{{Clark} {et~al.}(2009){Clark}, {Negueruela}, {Davies}, {Larionov},
  {Ritchie}, {Figer}, {Messineo}, {Crowther}, \& {Arkharov}}]{clark09}
{Clark}, J.~S., {Negueruela}, I., {Davies}, B., {et~al.} 2009, \aap, 498, 109

\bibitem[{{Cutri} \& {et al.}(2013)}]{cutri13}
{Cutri}, R.~M. \& {et al.} 2013, VizieR Online Data Catalog, 2328

\bibitem[{{Debattista} {et~al.}(2002){Debattista}, {Gerhard}, \&
  {Sevenster}}]{debattista02}
{Debattista}, V.~P., {Gerhard}, O., \& {Sevenster}, M.~N. 2002, \mnras, 334,
  355

\bibitem[{{Deguchi} {et~al.}(2004){Deguchi}, {Fujii}, {Glass}, {Imai}, {Ita},
  {Izumiura}, {Kameya}, {Miyazaki}, {Nakada}, \& {Nakashima}}]{deguchi04}
{Deguchi}, S., {Fujii}, T., {Glass}, I.~S., {et~al.} 2004, \pasj, 56, 765

\bibitem[{{Deguchi} {et~al.}(2000{\natexlab{a}}){Deguchi}, {Fujii}, {Izumiura},
  {Kameya}, {Nakada}, \& {Nakashima}}]{deguchi00b}
{Deguchi}, S., {Fujii}, T., {Izumiura}, H., {et~al.} 2000{\natexlab{a}}, \apjs,
  130, 351

\bibitem[{{Deguchi} {et~al.}(2000{\natexlab{b}}){Deguchi}, {Fujii}, {Izumiura},
  {Kameya}, {Nakada}, {Nakashima}, {Ootsubo}, \& {Ukita}}]{deguchi00a}
{Deguchi}, S., {Fujii}, T., {Izumiura}, H., {et~al.} 2000{\natexlab{b}}, \apjs,
  128, 571

\bibitem[{{Deguchi} {et~al.}(2000{\natexlab{c}}){Deguchi}, {Fujii}, {Izumiura},
  {Kameya}, {Nakada}, \& {Nakashima}}]{deguchi00}
{Deguchi}, S., {Fujii}, T., {Izumiura}, H., {et~al.} 2000{\natexlab{c}}, \apjs,
  130, 351

\bibitem[{{Deguchi} {et~al.}(2010){Deguchi}, {Nakashima}, {Zhang}, {Chong},
  {Koike}, \& {Kwok}}]{deguchi10}
{Deguchi}, S., {Nakashima}, J.-I., {Zhang}, Y., {et~al.} 2010, \pasj, 62, 391

\bibitem[{{Egan} {et~al.}(2003){Egan}, {Price}, {Kraemer}, {Mizuno}, {Carey},
  {Wright}, {Engelke}, {Cohen}, \& {Gugliotti}}]{egan03}
{Egan}, M.~P., {Price}, S.~D., {Kraemer}, K.~E., {et~al.} 2003, VizieR Online
  Data Catalog, 5114

\bibitem[{{Egan} {et~al.}(1999){Egan}, {Price}, {Moshir}, {Cohen}, {Tedesco},
  {Murdock}, {Zweil}, {Burdick}, {Bonito}, {Gugliotti}, \& {Duszlak}}]{egan99}
{Egan}, M.~P., {Price}, S.~D., {Moshir}, M., {et~al.} 1999, Air Force Research
  Lab. Technical Rep. AFRL-VS-TR-1999-1522 (1999)

\bibitem[{{Epchtein} {et~al.}(1994){Epchtein}, {de Batz}, {Copet}, {Fouque},
  {Lacombe}, {Le Bertre}, {Mamon}, {Rouan}, {Tiphene}, {Burton}, {Deul},
  {Habing}, {Boersenberger}, {Dennefeld}, {Omont}, {Renault},
  {Rocca-Volmerange}, {Kimeswenger}, {Appenzeller}, {Bender}, {Forveille},
  {Garzon}, {Hron}, {Persi}, {Ferrari-Toniolo}, \& {Vauglin}}]{epchtein94}
{Epchtein}, N., {de Batz}, B., {Copet}, E., {et~al.} 1994, \apss, 217, 3

\bibitem[{{Felli} {et~al.}(2002){Felli}, {Testi}, {Schuller}, \&
  {Omont}}]{felli02}
{Felli}, M., {Testi}, L., {Schuller}, F., \& {Omont}, A. 2002, \aap, 392, 971

\bibitem[{{Gutermuth} \& {Heyer}(2015)}]{gutermuth15}
{Gutermuth}, R.~A. \& {Heyer}, M. 2015, \aj, 149, 64

\bibitem[{{Habing} {et~al.}(2006){Habing}, {Sevenster}, {Messineo}, {van de
  Ven}, \& {Kuijken}}]{habing06}
{Habing}, H.~J., {Sevenster}, M.~N., {Messineo}, M., {van de Ven}, G., \&
  {Kuijken}, K. 2006, \aap, 458, 151

\bibitem[{{Halpern} \& {Gotthelf}(2007)}]{halpern07}
{Halpern}, J.~P. \& {Gotthelf}, E.~V. 2007, \apj, 669, 579

\bibitem[{{Iben} \& {Renzini}(1983)}]{iben83}
{Iben}, I. \& {Renzini}, A. 1983, \araa, 21, 271

\bibitem[{{Ishihara} {et~al.}(2011){Ishihara}, {Kaneda}, {Onaka}, {Ita},
  {Matsuura}, \& {Matsunaga}}]{ishihara11}
{Ishihara}, D., {Kaneda}, H., {Onaka}, T., {et~al.} 2011, \aap, 534, A79

\bibitem[{{Ishihara} {et~al.}(2010){Ishihara}, {Onaka}, {Kataza}, {Salama},
  {Alfageme}, {Cassatella}, {Cox}, {Garc{\'{\i}}a-Lario}, {Stephenson},
  {Cohen}, {Fujishiro}, {Fujiwara}, {Hasegawa}, {Ita}, {Kim}, {Matsuhara},
  {Murakami}, {M{\"u}ller}, {Nakagawa}, {Ohyama}, {Oyabu}, {Pyo}, {Sakon},
  {Shibai}, {Takita}, {Tanab{\'e}}, {Uemizu}, {Ueno}, {Usui}, {Wada},
  {Watarai}, {Yamamura}, \& {Yamauchi}}]{ishihara10}
{Ishihara}, D., {Onaka}, T., {Kataza}, H., {et~al.} 2010, \aap, 514, A1

\bibitem[{{Izumiura} {et~al.}(1999){Izumiura}, {Deguchi}, {Fujii}, {Kameya},
  {Matsumoto}, {Nakada}, {Ootsubo}, \& {Ukita}}]{izumiura99}
{Izumiura}, H., {Deguchi}, S., {Fujii}, T., {et~al.} 1999, \apjs, 125, 257

\bibitem[{{Jura} \& {Kleinmann}(1990)}]{jura90}
{Jura}, M. \& {Kleinmann}, S.~G. 1990, \apjs, 73, 769

\bibitem[{{Kwok} {et~al.}(1997){Kwok}, {Volk}, \& {Bidelman}}]{kwok97}
{Kwok}, S., {Volk}, K., \& {Bidelman}, W.~P. 1997, \apjs, 112, 557

\bibitem[{{Le Bertre} \& {Nyman}(1990)}]{lebertre90}
{Le Bertre}, T. \& {Nyman}, L.-A. 1990, \aap, 233, 477

\bibitem[{{Lindqvist} {et~al.}(1992){Lindqvist}, {Winnberg}, {Habing}, \&
  {Matthews}}]{lindqvist92}
{Lindqvist}, M., {Winnberg}, A., {Habing}, H.~J., \& {Matthews}, H.~E. 1992,
  \aaps, 92, 43

\bibitem[{{L{\'o}pez-Corredoira} {et~al.}(1999){L{\'o}pez-Corredoira},
  {Garz{\'o}n}, {Beckman}, {Mahoney}, {Hammersley}, \&
  {Calbet}}]{lopez-corredoira99}
{L{\'o}pez-Corredoira}, M., {Garz{\'o}n}, F., {Beckman}, J.~E., {et~al.} 1999,
  \aj, 118, 381

\bibitem[{{Lucas} {et~al.}(2008){Lucas}, {Hoare}, {Longmore}, {Schr{\"o}der},
  {Davis}, {Adamson}, {Bandyopadhyay}, {de Grijs}, {Smith}, {Gosling},
  {Mitchison}, {G{\'a}sp{\'a}r}, {Coe}, {Tamura}, {Parker}, {Irwin}, {Hambly},
  {Bryant}, {Collins}, {Cross}, {Evans}, {Gonzalez-Solares}, {Hodgkin},
  {Lewis}, {Read}, {Riello}, {Sutorius}, {Lawrence}, {Drew}, {Dye}, \&
  {Thompson}}]{lucas08}
{Lucas}, P.~W., {Hoare}, M.~G., {Longmore}, A., {et~al.} 2008, \mnras, 391, 136

\bibitem[{{Messineo} {et~al.}(2004){Messineo}, {Habing}, {Menten}, {Omont}, \&
  {Sjouwerman}}]{messineo04}
{Messineo}, M., {Habing}, H.~J., {Menten}, K.~M., {Omont}, A., \& {Sjouwerman},
  L.~O. 2004, \aap, 418, 103

\bibitem[{{Messineo} {et~al.}(2005){Messineo}, {Habing}, {Menten}, {Omont},
  {Sjouwerman}, \& {Bertoldi}}]{messineo05}
{Messineo}, M., {Habing}, H.~J., {Menten}, K.~M., {et~al.} 2005, \aap, 435, 575

\bibitem[{{Messineo} {et~al.}(2002){Messineo}, {Habing}, {Sjouwerman}, {Omont},
  \& {Menten}}]{messineo02}
{Messineo}, M., {Habing}, H.~J., {Sjouwerman}, L.~O., {Omont}, A., \& {Menten},
  K.~M. 2002, \aap, 393, 115

\bibitem[{{Messineo} {et~al.}(2016){Messineo}, {Zhu}, {Menten}, {Ivanov},
  {Figer}, {Kudritzki}, \& {Chen}}]{messineo16}
{Messineo}, M., {Zhu}, Q., {Menten}, K.~M., {et~al.} 2016, \apjl, 822, L5

\bibitem[{{Messineo} {et~al.}(2017){Messineo}, {Zhu}, {Menten}, {Ivanov},
  {Figer}, {Kudritzki}, \& {Chen}}]{messineo17}
{Messineo}, M., {Zhu}, Q., {Menten}, K.~M., {et~al.} 2017, \apj, 836, 65

\bibitem[{{Nagata} {et~al.}(1993){Nagata}, {Hyland}, {Straw}, {Sato}, \&
  {Kawara}}]{nagata93}
{Nagata}, T., {Hyland}, A.~R., {Straw}, S.~M., {Sato}, S., \& {Kawara}, K.
  1993, \apj, 406, 501

\bibitem[{{Negueruela} {et~al.}(2011){Negueruela},
  {Gonz{\'a}lez-Fern{\'a}ndez}, {Marco}, \& {Clark}}]{negueruela11}
{Negueruela}, I., {Gonz{\'a}lez-Fern{\'a}ndez}, C., {Marco}, A., \& {Clark},
  J.~S. 2011, \aap, 528, A59

\bibitem[{{Negueruela} {et~al.}(2010){Negueruela},
  {Gonz{\'a}lez-Fern{\'a}ndez}, {Marco}, {Clark}, \&
  {Mart{\'{\i}}nez-N{\'u}{\~n}ez}}]{negueruela10}
{Negueruela}, I., {Gonz{\'a}lez-Fern{\'a}ndez}, C., {Marco}, A., {Clark},
  J.~S., \& {Mart{\'{\i}}nez-N{\'u}{\~n}ez}, S. 2010, \aap, 513, A74

\bibitem[{{Omont} {et~al.}(2003){Omont}, {Gilmore}, {Alard}, {Aracil},
  {August}, {Baliyan}, {Beaulieu}, {B{\'e}gon}, {Bertou}, {Blommaert},
  {Borsenberger}, {Burgdorf}, {Caillaud}, {Cesarsky}, {Chitre}, {Copet}, {de
  Batz}, {Egan}, {Egret}, {Epchtein}, {Felli}, {Fouqu{\'e}}, {Ganesh},
  {Genzel}, {Glass}, {Gredel}, {Groenewegen}, {Guglielmo}, {Habing},
  {Hennebelle}, {Jiang}, {Joshi}, {Kimeswenger}, {Messineo},
  {Miville-Desch{\^e}nes}, {Moneti}, {Morris}, {Ojha}, {Ortiz}, {Ott},
  {Parthasarathy}, {P{\'e}rault}, {Price}, {Robin}, {Schultheis}, {Schuller},
  {Simon}, {Soive}, {Testi}, {Teyssier}, {Tiph{\`e}ne}, {Unavane}, {van Loon},
  \& {Wyse}}]{omont03}
{Omont}, A., {Gilmore}, G.~F., {Alard}, C., {et~al.} 2003, \aap, 403, 975

\bibitem[{{Ortiz} {et~al.}(2005){Ortiz}, {Lorenz-Martins}, {Maciel}, \&
  {Rangel}}]{ortiz05}
{Ortiz}, R., {Lorenz-Martins}, S., {Maciel}, W.~J., \& {Rangel}, E.~M. 2005,
  \aap, 431, 565

\bibitem[{{Reid} {et~al.}(2009){Reid}, {Menten}, {Zheng}, {Brunthaler},
  {Moscadelli}, {Xu}, {Zhang}, {Sato}, {Honma}, {Hirota}, {Hachisuka}, {Choi},
  {Moellenbrock}, \& {Bartkiewicz}}]{reid09}
{Reid}, M.~J., {Menten}, K.~M., {Zheng}, X.~W., {et~al.} 2009, \apj, 700, 137

\bibitem[{{Schuller} {et~al.}(2003){Schuller}, {Ganesh}, {Messineo}, {Moneti},
  {Blommaert}, {Alard}, {Aracil}, {Miville-Desch{\^e}nes}, {Omont},
  {Schultheis}, {Simon}, {Soive}, \& {Testi}}]{schuller03}
{Schuller}, F., {Ganesh}, S., {Messineo}, M., {et~al.} 2003, \aap, 403, 955

\bibitem[{{Sevenster}(1999)}]{sevenster99}
{Sevenster}, M.~N. 1999, \mnras, 310, 629

\bibitem[{{Sevenster}(2002)}]{sevenster02}
{Sevenster}, M.~N. 2002, \aj, 123, 2772

\bibitem[{{Sevenster} {et~al.}(1997{\natexlab{a}}){Sevenster}, {Chapman},
  {Habing}, {Killeen}, \& {Lindqvist}}]{sevenster97a}
{Sevenster}, M.~N., {Chapman}, J.~M., {Habing}, H.~J., {Killeen}, N.~E.~B., \&
  {Lindqvist}, M. 1997{\natexlab{a}}, \aaps, 122

\bibitem[{{Sevenster} {et~al.}(1997{\natexlab{b}}){Sevenster}, {Chapman},
  {Habing}, {Killeen}, \& {Lindqvist}}]{sevenster97b}
{Sevenster}, M.~N., {Chapman}, J.~M., {Habing}, H.~J., {Killeen}, N.~E.~B., \&
  {Lindqvist}, M. 1997{\natexlab{b}}, \aaps, 124

\bibitem[{{Sevenster} {et~al.}(2001){Sevenster}, {van Langevelde}, {Moody},
  {Chapman}, {Habing}, \& {Killeen}}]{sevenster01}
{Sevenster}, M.~N., {van Langevelde}, H.~J., {Moody}, R.~A., {et~al.} 2001,
  \aap, 366, 481

\bibitem[{{Sjouwerman} {et~al.}(1998){Sjouwerman}, {van Langevelde},
  {Winnberg}, \& {Habing}}]{sjouwerman98}
{Sjouwerman}, L.~O., {van Langevelde}, H.~J., {Winnberg}, A., \& {Habing},
  H.~J. 1998, \aaps, 128, 35

\bibitem[{{Skrutskie} {et~al.}(2006){Skrutskie}, {Cutri}, {Stiening},
  {Weinberg}, {Schneider}, {Carpenter}, {Beichman}, {Capps}, {Chester},
  {Elias}, {Huchra}, {Liebert}, {Lonsdale}, {Monet}, {Price}, {Seitzer},
  {Jarrett}, {Kirkpatrick}, {Gizis}, {Howard}, {Evans}, {Fowler}, {Fullmer},
  {Hurt}, {Light}, {Kopan}, {Marsh}, {McCallon}, {Tam}, {Van Dyk}, \&
  {Wheelock}}]{twomass}
{Skrutskie}, M.~F., {Cutri}, R.~M., {Stiening}, R., {et~al.} 2006, \aj, 131,
  1163

\bibitem[{{Verheyen} {et~al.}(2012){Verheyen}, {Messineo}, \&
  {Menten}}]{verheyen12}
{Verheyen}, L., {Messineo}, M., \& {Menten}, K.~M. 2012, \aap, 541, A36

\bibitem[{{Wright} {et~al.}(2010){Wright}, {Eisenhardt}, {Mainzer}, {Ressler},
  {Cutri}, {Jarrett}, {Kirkpatrick}, {Padgett}, {McMillan}, {Skrutskie},
  {Stanford}, {Cohen}, {Walker}, {Mather}, {Leisawitz}, {Gautier}, {McLean},
  {Benford}, {Lonsdale}, {Blain}, {Mendez}, {Irace}, {Duval}, {Liu}, {Royer},
  {Heinrichsen}, {Howard}, {Shannon}, {Kendall}, {Walsh}, {Larsen}, {Cardon},
  {Schick}, {Schwalm}, {Abid}, {Fabinsky}, {Naes}, \& {Tsai}}]{wright10}
{Wright}, E.~L., {Eisenhardt}, P.~R.~M., {Mainzer}, A.~K., {et~al.} 2010, \aj,
  140, 1868

\bibitem[{{Yamamura} {et~al.}(2010){Yamamura}, {Makiuti}, {Ikeda}, {Fukuda},
  {Oyabu}, {Koga}, \& {White}}]{yamamura10}
{Yamamura}, I., {Makiuti}, S., {Ikeda}, N., {et~al.} 2010, VizieR Online Data
  Catalog, 2298

\end{thebibliography}

\begin{appendix} 

\section{Spectra}

The spectra of targets with detected SiO maser are shown in Fig. \ref{spectra}.

\begin{figure*}
\begin{center}
\resizebox{1\hsize}{!}{\includegraphics[angle=0]{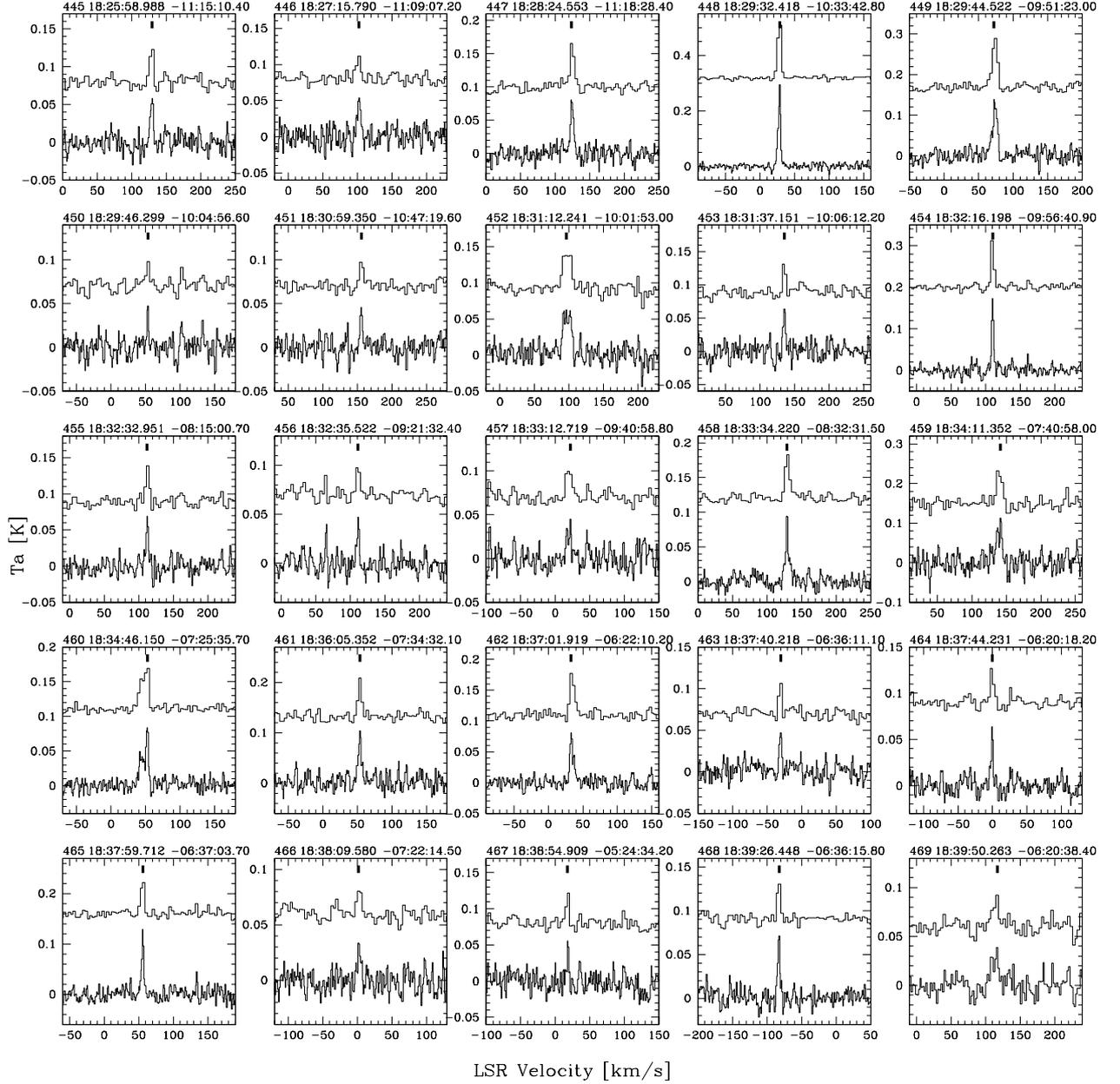}}
\caption{\label{spectra} IRAM spectra of 86 GHz SiO maser lines. Each panel shows at the bottom the spectrum obtained with the autocorrelator (1.1 \kms), 
and at the top that obtained with the filter bank (3.5 \kms, offset from zero).
The line-of-sight velocity listed in Table 1 is indicated with a small vertical bar at the top of each panel. 
The conversion factor between antenna temperatures and flux densities  is 6.2 Jy/K. } 
\end{center}
\end{figure*}

\addtocounter{figure}{-1}
\begin{figure*}
\begin{center}
\resizebox{1\hsize}{!}{\includegraphics[angle=0]{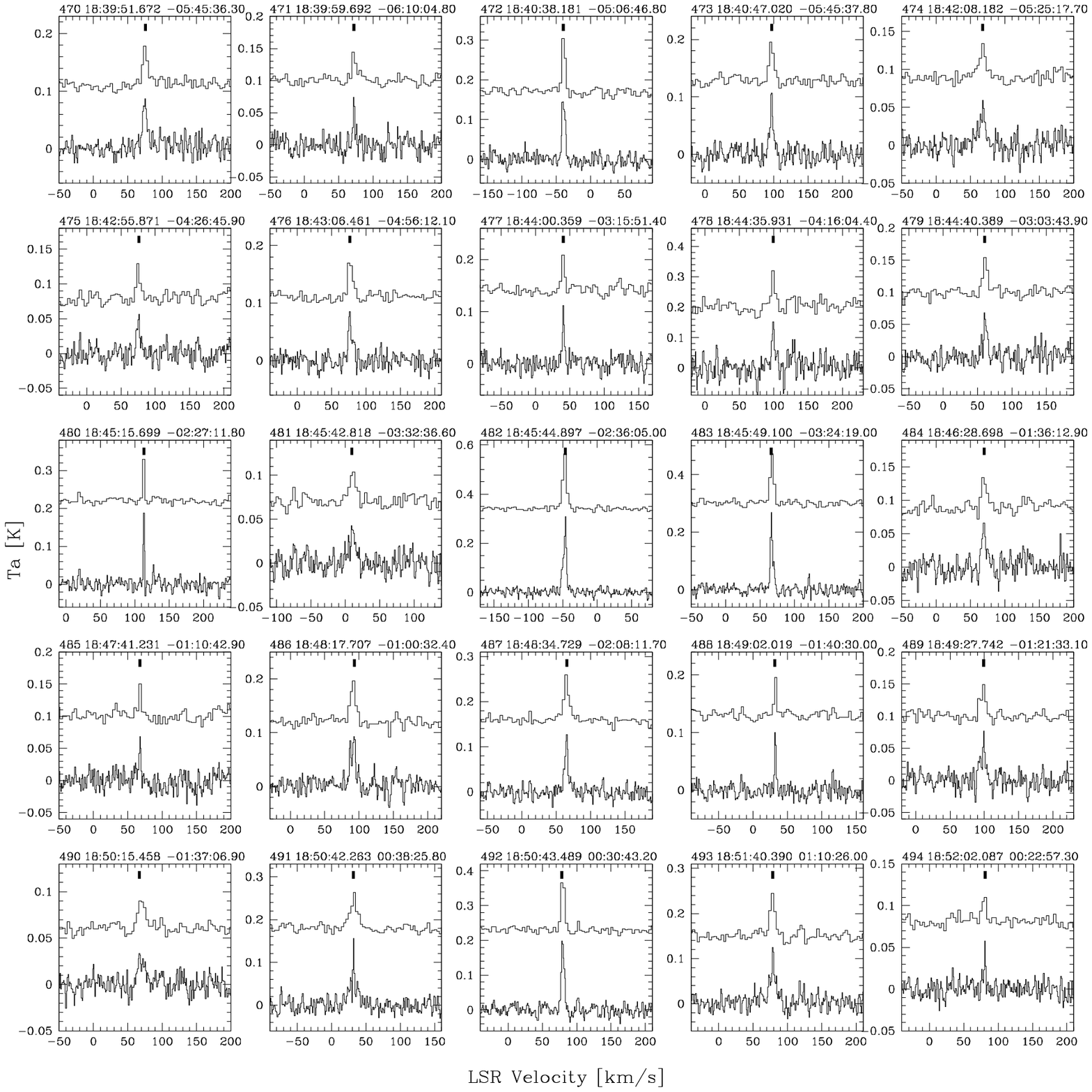}}
\caption{  Continuation of Fig.\ \ref{spectra}. } 
\end{center}
\end{figure*}

\addtocounter{figure}{-1}
\begin{figure*}
\begin{center}
\resizebox{1\hsize}{!}{\includegraphics[angle=0]{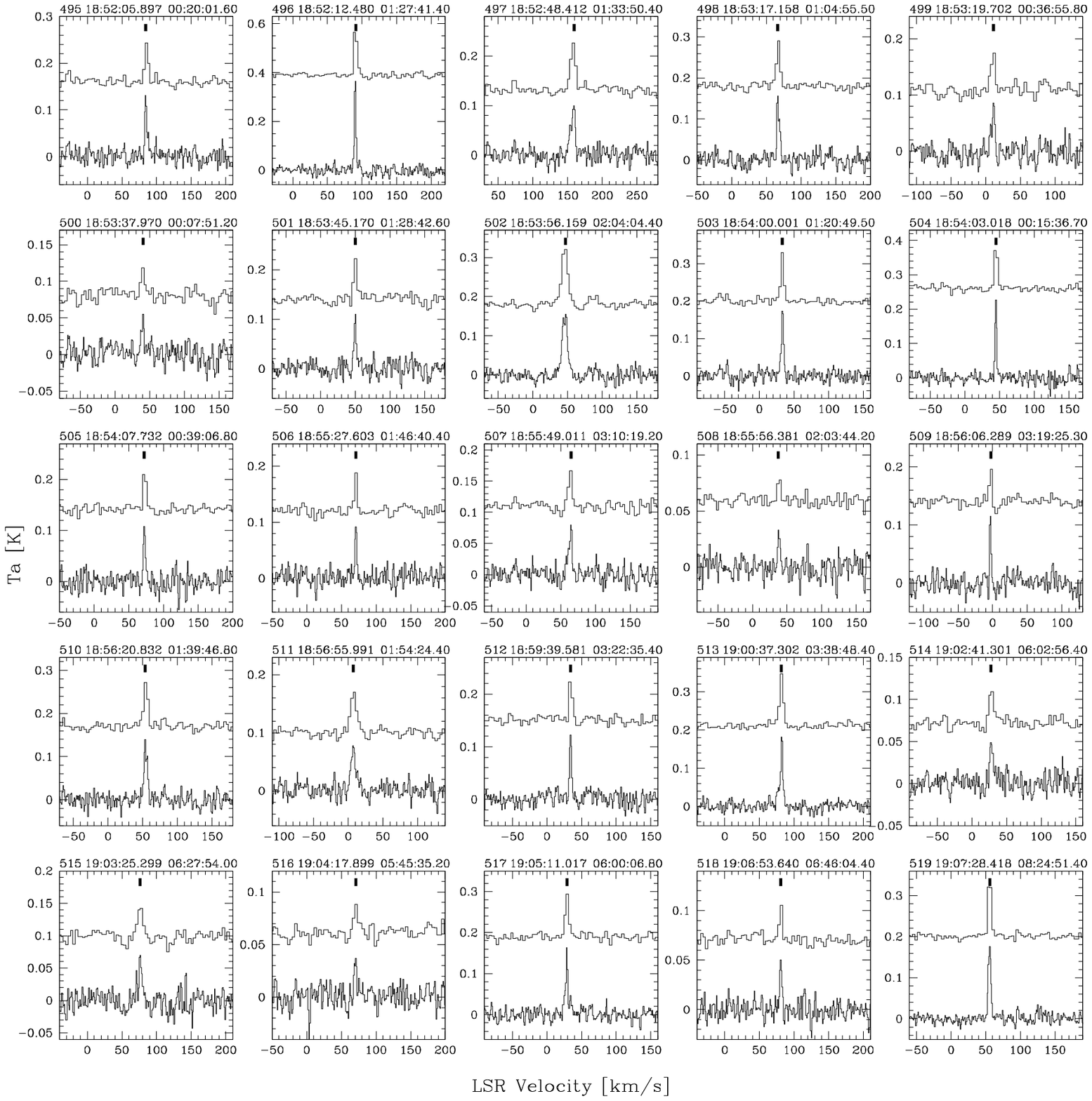}}
\caption{Continuation of Fig.\ \ref{spectra}. } 
\end{center}
\end{figure*}

\addtocounter{figure}{-1}
\begin{figure*}
\begin{center}
\resizebox{1\hsize}{!}{\includegraphics[angle=0]{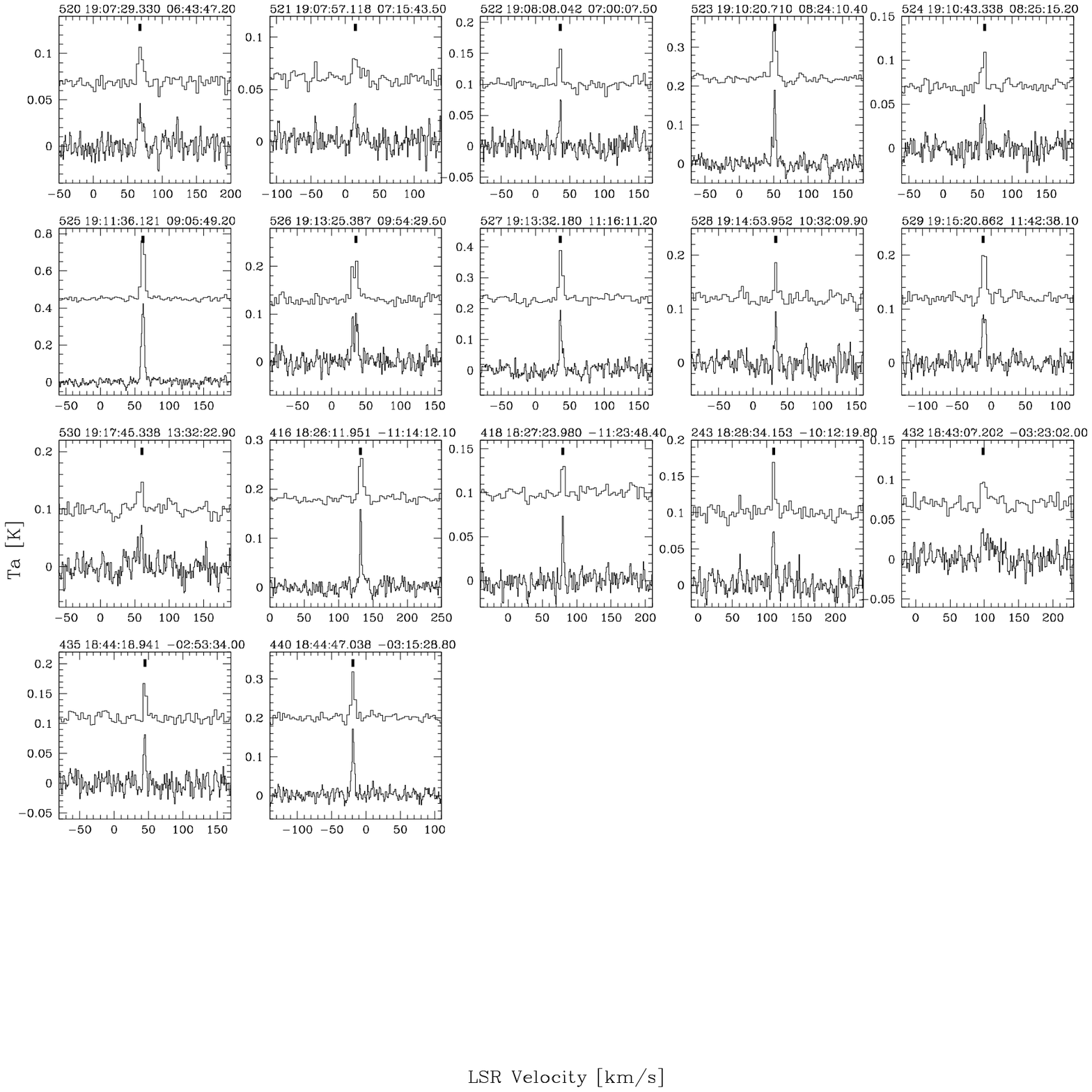}}
\caption{Continuation of Fig.\ \ref{spectra}. } 
\end{center}
\end{figure*}

\section{Notes to the infrared  catalogs}
\label{Kmatches}

2MASS matches were provided
by the two catalogs WISE and GLIMPSE. 
We generally adopted the AllWISE Data Release, which has an improved astrometry, but,
 the number of targets  was in only  the earlier WISE All-Sky Data Release.
When 2MASS matches were missed by GLIMPSE (mostly due to saturation effects), 
we inspected the images  (GLIMPSE, WISE, and 2MASS) and  retained the WISE matches.
We inspected the images  and retained the GLIMPSE values when 2MASS matches had been missed by WISE (mostly due to saturation effects and crowding).
We retained the 2MASS matches provided by GLIMPSE for targets \#58, \#78, \#320, \#342,  and \#493.
One near-infrared star (\#39)  is resolved by UKIDSS into two similarly bright stars. 
UKIDSS data were used to revise the positions and magnitudes of \#58,  \#320,  and 
\#413 because they were blended or missed in 2MASS. 
We assigned UKIDSS magnitudes for four entries without 2MASS data (\#224, \#298, \#423, and \#536). 

In \citet{messineo04} 2MASS, ISOGAL, and MSX (v1.2) counterparts 
have already been provided for the 444 stars in \citet{messineo02}.
The MSX magnitudes tabulated in this paper differ from those in \citet{messineo04} because 
 there we used an earlier version of the MSX catalog.
We compared the \Ks\ measurements listed by \citet{messineo04} with those
in Table \ref{table.magnitudes} and found only five 2MASS mismatches 
(targets   \#77, \#90, \#227,  \#347,  and  \#424) and the updated seven   \Ks\ values 
with measurements from the UKIDSS catalog.
Masers \#77  and \#78  were detected in the same beam. 
The maser with the stronger flux 
was associated with the pointed star ISOGAL$-J$174618.9-284439. 
The fainter maser \#77 had been  included in  \citet{messineo04} and had been   
associated with the only  star detected at \#7 \um\ by ISOGAL about 12\farcs5  away. 
The GLIMPSE catalog allows us to revise this association with the closer  
GLIMPSE star G000.2444$-$00.0294 (5\farcs2 away, [8.0]=4.93 mag and \Ks=7.78 mag).
Stars  \#21 and \#22 are  double detections within the same pointing, 
identified with two ISOGAL point sources.
The two masers \#64 and \#65  were both detected when pointing toward 
the mid-infrared source  ISOGAL$-J$174528.8-284734. 
Possible fainter stars ($\approx8$ mag) falling in the IRAM beam are detected in the 8 \um\ 
GLIMPSE images; however, they are below [8.0]= 8.0 mag, while the average brightness of our 
SiO masing stars is [8.0]= 4.2 mag  with a  $\sigma$=0.6 mag.

The WISE counterparts were searched within 10\arcsec,
and the closest was retained. We found 28 cases  of multiple stars within the search radius.
The closest match  was the brightest in the $W3$ band in all cases but \#110 and \#405.
The reddest \Ks$-W3$ match was not the closest  in the cases of
stars \#21 (9.4\arcsec\ away), \#35 (6.7\arcsec\ away), 
\#43 (9.7\arcsec\ away), \#110 (8.0\arcsec\ away), 
\#173 (6\farcs6 away)
\#207 (7\farcs8 away), \#228 (6\farcs1 away), 
\#289 (9\farcs5 away), \#324 (4\farcs4 away), 
\#354 (4\farcs7 away), \#405 (9\farcs1 away), 
\#479 (9\farcs8 away), \#482 (9\farcs8 away), 
\#502 (9\farcs6 away), \#545 (8\farcs3 away), 
and \#548 (8\farcs7 away).
We  searched for AKARI/FIS data at 65, 90, and 140 \um\ \citep{yamamura10}, but, unfortunately,
matches were available only for three targets (\#251, \#451, and \#537).

\section{Contamination RSGs}
\label{mbol}

To asses the nature of a few suspect broad and
multipeaked  maser lines, which  are reported in Sect. \ref{fwhm},
we estimated  luminosities and distances.
Extinction corrections are as in  \citet[][]{messineo05}.  
Bolometric  magnitudes were computed
by integrating the dereddened flux densities, $F_\nu(\nu)$ over frequency $\nu$ 
with linear interpolations and by extrapolating to $F_{\nu=0}=0$
and at the upper end  with a blackbody curve.
The low-frequency extrapolation is insignificant since it contains 
a negligible fraction of the total flux (3\% in average). 
When both DENIS and 2MASS datasets were available, we averaged the 
two estimates to attenuate the effect of variability;
the average difference being 0.036 mag with $\sigma=$ 0.31 mag.

In Table \ref{table.luminosity}, we  list \Mbol\ values of the brighest
stars. The list is useful because it contains probable RSG stars, but 
infrared spectroscopy is needed to firmly unveil the nature of those bright stars.
The Table collects the brighest stars from  
five computations with different assumptions on distances.

\begin{itemize}
\item[] {\it Group 1}:  Previously known RSGs included in the sample.
\item[] {\it Group 2}:  Stars located in the central 5\degr\ of longitude which are brighter than the AGB limit, 
                     $M_{\mathrm{bol}}<-7.2$ mag, when  a distance modulus of 14.5 mag  is assumed.
\item[] {\it Group 3}: Brighest stars  with longitude $>20^\circ$ for which kinematic near distances are used.
                     
\item[] {\it Group 4}: Brighest stars with longitude   $<20^\circ$ and classified as foreground to the bar and bulge in \citet{messineo05}.
                       For these stars, kinematic distances are assumed.

\item[] {\it Group 5}: List of all remaining targets with FWHM $>$ 9 \kms.
\end{itemize}

{\it Group 1 :} The sample contains two spectroscopically identified RSGs,  \#436 and \#444
\citep{jura90,messineo17}, and two photometrically identified RSGs, 
\#483 and \#535  \citep{clark09,negueruela10}. 
Only star \#483 (l=29\degr) is a maser detection; 
it has  a velocity \Vlsr=66.89 \kms, and  \Mbol=$-5.63$ mag.

{\it Group 2 :} For the 251 targets located within the inner 5\degr\  in longitude,
we initially assume a distance of 8.0 kpc. 
The bulk of targets have $M_{\mathrm{bol}}$ from $-4.5$
to $-6.0$ mag as expected for thermally pulsing AGB stars. 
A bright tail in the luminosity distribution  suggests that a few RSGs are 
included in the sample.
A number of 17 stars (7\% of targets within $5^\circ$)   are brighter than  $M_{\mathrm{bol}}=-7.2$ mag, 
which is the classical AGB limit \citep{iben83}. 
Ten of them are likely to be foreground stars, 
because their total extinction 
is lower than the average interstellar extinction of surrounding stars \citep{messineo05}. 
The remaining seven bright sources 
(\#31,  \#75, \#92, \#116, \#128, \#294, and \#310) 
with average $M_{\mathrm{bol}} < -7.2$ mag 
have total  extinction  very close to that measured from surrounding stars,
therefore, \Mbol\ does not depend on the choice of \Aks.
For  stars, \#31,  \#92, \#128, and \#294, infrared spectra 
are analyzed in the work by \citet[][]{messineo17}. 
Stars \#31 (\Mbol(2MASS)=$-7.47$ and \Mbol(DENIS)=$-7.17$ mag) and 
\#128 ( \Mbol(2MASS)=$-7.94$ and \Mbol(DENIS)= $-6.49$ mag) 
have some water absorption which  suggests AGB stars. 
Stars \#92 and \#294 have a low water content and broad CO bands that are 
typical of RSGs. Their \Mbol(2MASS) are $-8.22$  and $-7.76$ mag.  
Star \#92  (\Vlsr=119 \kms) is already classified as 
a RSG by \citet{nagata93} and \citet{messineo17}.

{\it Group 3 :} Using kinematic distances \citep{reid09}, their  \Mbol\ values do not exceed the AGB limit.
We note that in {\it Group 2} most of the stars  have $M_{\mathrm{bol}}$ from $-4.5$
to $-6.0$ mag.
We are able to locate  eight bright stars at $l>20^\circ$ with \Mbol $< -6.0$ mag 
(i.e., 7\% of detections at $l>20^\circ$).

{\it Group 4 :} Stars defined as foreground by \citet{messineo05} may include some RSGs.
Their \Aks\ values is smaller than that of the average of surrounding stars.
Indeed, \Mbol\ estimated for stars \#15 and \#129 approach the AGB limit.

{\it Group 5 :} Most of the remaining stars with  FWHMs $>9$ \kms\ 
have \Mbol\ values larger than $-5.5$ mag.
We estimate that the RSG contamination is $\la 7$\% (Table \ref{table.luminosity}).
With our signal-to-noise it is not possible to use the FWHM of maser lines to establish
luminosity classes; however, candidate RSGs have in average broader lines.
For example, at $l>20^\circ$ 15\% of the detections have broad FWHMs ($>9$ \kms),
but only 7\% of the detections are candidate RSGs (25\%  of which have FWHMs $>9$ \kms).

\newpage
\begin{table*}\renewcommand{\arraystretch}{0.8}
\caption{\label{table.luminosity}  Luminosities of bright targets  and of targets with broad maser lines.} 
{\tiny
\begin{tabular}{rrrrrrrrll}
\hline
ID    &  Long    & \Ak(field)&  \Ak(tot)&  \Mbol(kin)& \Mbol(14.5)& \Vlsr & FWHM &  Comments  & References\\
      &   [\degr]  &  [mag]  & [mag]    &     [mag]  &  [mag]     &[\kms]  & [\kms]   & \\ 
\hline
\hline
436    &  29.85   &   0.12   &   $..$0   &    $..$   &   $-$8.4   &    $..$   &    $..$   &                              M3I &(1)    \\
444    &  29.67   &   1.16   &   0.88   &    $..$   &   $-$7.9   &    $..$   &    $..$   &                               M2I &(2)  \\
483    &  29.26   &   1.32   &   1.40   &   $-$5.6   &   $-$7.1   &   66.9   &    4.3   &                                RSG$^a$ &(3)    \\
535    &  24.63   &   1.29   &   1.76   &    $..$   &   $-$5.0   &    $..$   &    $..$   &                               RSG$^a$ & (4) \\
\hline

 31    &  $-$0.44   &   1.80   &   1.72   &    $..$   &   $-$7.3   &  $-$24.4   &    9.6   &  \Mbol(14.5)$<$-$7.2$, long$<5^\circ$    \\
 75    &     0.77   &   1.36   &   1.20   &    $..$   &   $-$8.0   &  $-$38.1   &    3.3   &  \Mbol(14.5)$<$-$7.2$, long$<5^\circ$    \\
 92    &     0.11   &   1.66   &   1.41   &    $..$   &   $-$8.2   &    119.4   &    2.9   &  \Mbol(14.5)$<$-$7.2$, long$<5^\circ$, sRSG$^b$ & (5)    \\
116    &     2.49   &   1.75   &   1.40   &    $..$   &   $-$7.6   &     19.5   &    4.2   &  \Mbol(14.5)$<$-$7.2$, long$<5^\circ$    \\
128    &     2.70   &   1.51   &   1.60   &    $..$   &   $-$7.9   &  $-$16.4   &    4.5   &  \Mbol(14.5)$<$-$7.2$, long$<5^\circ$    \\
294    &  $-$1.00   &   2.23   &   2.37   &    $..$   &   $-$7.8   &      $..$   &    $..$   &  \Mbol(14.5)$<$-$7.2$, long$<5^\circ$, sRSG$^b$&  (5)  \\
310    &  $-$0.55   &   2.79   &   2.42   &    $..$   &   $-$7.4   &      $..$   &    $..$   &  \Mbol(14.5)$<$-$7.2$, long$<5^\circ$    \\
\hline
244    &  20.40   &   0.96   &   0.99   &   $-$6.2   &   $-$7.1   &   89.1   &    3.3   &   \Mbol(kin)$<-6.0$, long$>20^\circ$    \\
449    &  21.70   &   1.05   &   0.91   &   $-$6.3   &   $-$7.6   &   73.6   &    8.2   &   \Mbol(kin)$<-6.0$, long$>20^\circ$    \\
456    &  22.46   &   1.61   &   1.36   &   $-$6.6   &   $-$7.3   &  111.1   &    3.5   &   \Mbol(kin)$<-6.0$, long$>20^\circ$    \\
460    &  24.43   &   1.20   &   0.98   &   $-$6.2   &   $-$8.1   &   49.0   &   13.2   &   \Mbol(kin)$<-6.0$, long$>20^\circ$    \\
473    &  26.59   &   0.72   &   0.65   &   $-$6.1   &   $-$7.0   &   96.8   &    6.0   &   \Mbol(kin)$<-6.0$, long$>20^\circ$    \\
474    &  27.05   &   0.82   &   0.50   &   $-$6.4   &   $-$7.9   &   66.9   &   11.8   &   \Mbol(kin)$<-6.0$, long$>20^\circ$    \\
484    &  30.94   &   0.56   &   0.35   &   $-$6.3   &   $-$7.7   &   69.2   &    7.3   &   \Mbol(kin)$<-6.0$, long$>20^\circ$    \\
495    &  33.30   &   0.95   &   0.59   &   $-$6.1   &   $-$7.1   &   84.5   &    3.8   &   \Mbol(kin)$<-6.0$, long$>20^\circ$    \\
\hline
 15    &  $-$1.38   &   0.99   &   0.89   &   $-$6.9   &   $-$7.2   &   38.7   &    7.3   & \Mbol(kin)$<-6.0$, long$<20^\circ$, FG$^c$     \\
 23    &  $-$1.29   &   1.22   &   1.07   &   $-$6.8   &   $-$7.0   &   56.5   &    2.8   & \Mbol(kin)$<-6.0$, long$<20^\circ$, FG$^c$     \\
113    &   1.98   &   1.31   &   1.13   &   $-$6.7   &   $-$7.1   &   31.7   &   13.9   & \Mbol(kin)$<-6.0$, long$<20^\circ$, FG$^c$     \\
129    &   3.82   &   0.47   &   0.36   &   $-$7.1   &   $-$7.7   &   38.0   &   13.9   & \Mbol(kin)$<-6.0$, long$<20^\circ$, FG$^c$     \\
130    &   4.49   &   0.68   &   0.52   &   $-$6.4   &   $-$6.9   &   52.8   &    3.2   & \Mbol(kin)$<-6.0$, long$<20^\circ$, FG$^c$     \\
160    &   6.96   &   0.82   &   0.82   &   $-$6.2   &   $-$7.3   &   39.9   &    7.8   & \Mbol(kin)$<-6.0$, long$<20^\circ$, FG$^c$     \\
\hline
 71    &   0.32   &   1.84   &   2.12   &    $..$   &   $-$5.3   &   92.5   &    9.0   &                  remaining FWHM $ >9$    \\
 87    &   1.24   &   1.21   &   1.24   &    $..$   &   $-$7.2   &  $-$15.3   &    9.2   &                remaining FWHM $ >9$    \\
117    &   2.01   &   1.85   &   3.25   &    $..$   &   $-$5.6   &   22.2   &   16.3   &                  remaining FWHM $ >9$    \\
135    &   3.81   &   1.20   &   1.46   &    $..$   &   $-$5.6   &  207.1   &   11.9   &                  remaining FWHM $ >9$    \\
173    &   8.66   &   1.61   &   1.20   &    $..$   &   $-$7.5   &   $-$1.1   &   11.5   &                remaining FWHM $ >9$    \\
190    &  10.99   &   1.04   &   1.10   &    $..$   &   $-$6.4   &   58.1   &    9.1   &                  remaining FWHM $ >9$    \\
203    &  14.31   &   0.91   &   3.14   &    $..$   &   $-$4.0   &   81.1   &   12.7   &                  remaining FWHM $ >9$    \\
223    &  17.80   &   1.02   &   1.87   &   $..$   &   $-$5.2   &  143.9   &   13.4   &                   remaining FWHM $ >9$    \\
232    &  17.92   &   1.10   &   1.45   &   $..$   &   $-$5.4   &   94.3   &    9.8   &                   remaining FWHM $ >9$    \\
452    &  21.71   &   1.11   &   1.54   &   $-$5.7   &   $-$6.6   &   97.1   &   13.3   &                 remaining FWHM $ >9$    \\
457    &  22.25   &   1.51   &   1.28   &   $-$4.3   &   $-$7.5   &   19.8   &    9.1   &                 remaining FWHM $ >9$    \\
459    &  24.13   &   1.05   &   1.66   &   $-$5.5   &   $-$5.6   &  139.5   &   10.4   &                 remaining FWHM $ >9$    \\
469    &  25.97   &   0.86   &   1.70   &   $-$4.3   &   $-$4.9   &  113.4   &   12.7   &                 remaining FWHM $ >9$    \\
481    &  29.12   &   0.73   &   0.31   &   $-$3.2   &   $-$7.8   &   10.2   &   11.7   &                 remaining FWHM $ >9$    \\
486    &  31.67   &   1.75   &   5.20   &   $-$4.2   &   $-$5.2   &   91.0   &   10.6   &                 remaining FWHM $ >9$    \\
489    &  31.50   &   1.43   &   2.68   &   $-$4.7   &   $-$5.4   &   98.0   &    9.8   &                 remaining FWHM $ >9$    \\
490    &  31.36   &   1.55   &   2.27   &   $-$4.0   &   $-$5.5   &   69.8   &   12.4   &                 remaining FWHM $ >9$    \\
491    &  33.42   &   1.11   &   1.09   &   $-$4.2   &   $-$7.0   &   31.5   &   14.6   &                 remaining FWHM $ >9$    \\
493    &  34.00   &   1.04   &   1.38   &   $-$5.0   &   $-$6.1   &   78.9   &   11.1   &                 remaining FWHM $ >9$    \\
502    &  35.06   &   2.02   &   2.02   &   $-$5.4   &   $-$7.5   &   45.8   &    9.1   &                 remaining FWHM $ >9$    \\
511    &  35.25   &   1.47   &   2.49   &   $-$0.9   &   $-$6.2   &    7.3   &    9.2   &                 remaining FWHM $ >9$    \\
526    &  44.24   &   0.94   &   1.39   &   $-$3.2   &   $-$5.7   &   34.0   &    9.7   &                 remaining FWHM $ >9$    \\
530    &  47.95   &   0.71   &   1.88   &   $-$4.0   &   $-$4.8   &   58.0   &   10.4   &                 remaining FWHM $ >9$    \\
432    &  28.97   &   0.97   &   1.17   &   $-$5.4   &   $-$6.2   &  103.1   &   19.9   &                 remaining FWHM $ >9$    \\
\hline
\end{tabular}
}
\begin{list}{}
\item {\bf Notes.} Identification numbers (ID) are followed by longitudes (Long), median extinction of surrounding stars 
\Ak(field), total extinction of the targets \Ak(tot) \citep{messineo05}, \Mbol\ values obtained by adopting 
the near kinematic distances, \Mbol(kin), \Mbol\ values obtained by adopting a distance  modulus of 14.5 mag,
radial velocities, \Vlsr, and the maser line FWHMs.\\
{\bf References.} 1=\citet{jura90}  ;  2= \citet{messineo16}; 3=\citet{clark09}; 4=\citet{negueruela10}; 5= this work.~\\
$^a$ RSG = Star photometrically reported as a probable RSG in previous studies.~\\
$^b$ sRSG = Stars confirmed  as RSGs in this work with the spectral indexes of \citet{messineo17}.~\\ 
$^c$ FG = Stars classified as likely foreground by \citet{messineo05}, their  \Aks\ values are smaller than the average \Aks\ of surrounding stars.\\
\end{list}
\end{table*}

\end{appendix}

\addtocounter{table}{+2} 
\begin{sidewaystable*} \renewcommand{\arraystretch}{0.8}
\vspace*{+18.5cm} %;;;two columns
\begin{center}
{\tiny
\caption{\label{table.magnitudes}  Infrared measurements of the targeted stars$^{**}$}. 

\begin{tabular}{@{\extracolsep{-.10 in}} r|rrrrr|rrrrr|rrrr|rrrr|rrrr|r|rr|rlr}
\hline 
 &    \multicolumn{5}{c}{\rm 2MASS} &    \multicolumn{5}{c}{\rm DENIS-ISOGAL}   &   \multicolumn{4}{c}{\rm GLIMPSE}   & \multicolumn{4}{c}{\rm MSX}&  \multicolumn{4}{c}{\rm WISE}  & \multicolumn{1}{c}{\rm MIPS} & \multicolumn{2}{c}{\rm AKARI}& Nstar$^+$  \\ 
\hline 
 {\rm ID} & RA(J2000) & DEC(J2000)    &   {\it J} & {\it H} & { $K_S$}  &
 {\it I} & {\it J} & { $K_S$}  & {\rm [7]} & {\rm [15]} &
  {\rm [3.6]} & {\rm [4.5]} & {\rm [5.8]} & {\rm [8.0]} &
 {\it A}  & {\it C}  &{\it D}  &{\it E}  &
 {\it W1} &{\it W2}  & {\it W3} &  {\it W4} & {\rm [24]} & {\rm [9]} & {\rm [18]} &\\ 
\hline 
  &  &    &    1.2 &  1.6 & 2.2  &
   0.8 &  1.2 & 2.2  & 7 & 15 &
  3.6 & 4.5 & 5.8 & 8.0 &
 8.3  & 12.1  &14.6  &21.3  &
 3.4 &4.6  & 11.6 &  22.1 & 23.7 & 9.0 & 18.0 & \\ 
\hline 
&  &   &    {\rm [mag]}   & {\rm [mag]} & {\rm [mag]} & 
 {\rm [mag]}  & {\rm [mag]} & {\rm [mag]} & {\rm [mag]} & {\rm [mag]}
& {\rm [mag]} & {\rm [mag]}  & {\rm [mag]}&{\rm [mag]}&{\rm [mag]}&{\rm [mag]} &{\rm [mag]}&{\rm [mag]}&{\rm [mag]}& {\rm [mag]}& {\rm [mag]}& {\rm [mag]}&{\rm [mag]}& {\rm [mag]}&{\rm [mag]}& \\ 
\hline 
                             1            &  17 31 40.98& $-32  $  03 55.94  &   12.53              &     9.66              &     8.10              & $..$ & 12.77 &  8.28 &  5.16 &  3.48 &  6.93 &  6.98 &  5.74 &  5.16 &  5.30 & 98.98 & 98.98 & 98.98 & 94.94 & 94.94 &  4.74 &  3.33 &   2.58 &  96.96 &   2.94 & 111 &   \\
                             2            &  17 36 42.18& $-30  $  59 11.74  &   14.67              &    10.42              &     8.09              & $..$ & 14.92 &  8.16 &  4.89 &  3.06 & 88.88 & 88.88 & 88.88 & 88.88 &  3.80 &  2.87 &  2.27 &  1.55 &  6.45 &  5.03 &  3.42 &  1.76 &   1.92 &   3.77 &   2.71 & 111 &   \\
                             3            &  17 37 07.29& $-31  $  21 31.29  &   12.42              &     9.30              &     7.60              & 94.94 & 12.22 &  7.53 &  4.99 &  3.55 & 88.88 & 88.88 & 88.88 & 88.88 &  4.86 & 98.98 & 98.98 & 98.98 &  6.19 &  5.25 &  3.97 &  2.75 &   2.97 &   4.64 &   3.23 & 111 &   \\
                             4            &  17 37 29.35& $-31  $  17 16.62  & $(15.41           )$ &    10.59              &     8.08              & $..$ & 15.01 &  7.97 &  4.31 &  3.44 & 88.88 & 88.88 & 88.88 & 88.88 &  4.84 & 98.98 &  3.17 & 98.98 &  6.55 &  5.50 &  4.43 &  3.17 &   2.68 &   4.68 &   3.57 & 111 &   \\
                             5            &  17 38 11.78& $-31  $  46 27.03  &   11.62              &     8.70              &     7.02              & $..$ & 11.78 &  7.02 & 96.96 & 96.96 & 88.88 & 88.88 & 88.88 & 88.88 &  3.38 &  2.23 &  2.16 & 98.98 &  5.15 &  4.19 &  2.65 &  1.50 &   1.63 &   3.69 &   2.19 & 111 &   \\
                             6            &  17 38 12.49& $-29  $  39 38.53  &   11.73              &     9.69              &     8.29              & 17.36 & 11.77 &  8.06 &  5.26 &  2.89 & 88.88 & 88.88 & 88.88 & 88.88 &  4.17 &  3.09 &  3.13 & 98.98 &  8.75 &  8.71 &  8.22 & 98.98 &  96.96 &   4.35 &   3.49 & 111 &   \\
                             7            &  17 38 17.07& $-29  $  42 32.42  &    7.75              &     6.22              &     5.38              & 12.65 & $..$ &  6.42 &  4.63 &  3.02 &  4.62 &  4.86 &  4.46 &  4.21 & 97.97 & 97.97 & 97.97 & 97.97 &  4.59 &  4.35 &  3.82 &  3.05 &   3.33 &   4.18 &  96.96 & 121 &   \\
                             8            &  17 38 29.01& $-31  $  26 17.49  &   11.01              &     7.94              &     6.29              & $..$ & 10.68 &  6.85 &  4.04 &  2.93 & $..$ &  5.28 &  4.48 &  4.38 &  4.24 &  3.08 &  2.78 & 98.98 & 94.94 & 94.94 &  3.70 &  2.48 &   2.89 &   3.85 &   2.37 & 111 &   \\
                             9            &  17 38 32.50& $-31  $  20 42.75  & $(14.14           )$ &     9.99              &     7.75              & $..$ & 13.94 &  7.67 &  4.74 &  3.44 & 88.88 & 88.88 & 88.88 & 88.88 &  4.77 &  3.48 &  3.47 & 98.98 &  5.63 &  4.88 &  4.09 &  3.10 &   3.23 &  96.96 &   3.51 & 111 &   \\
                            10            &  17 38 35.69& $-29  $  36 37.21  &   12.58              &     9.80              &     8.12              & 19.15 & 11.21 &  7.33 &  4.59 & $..$ & 88.88 & 88.88 & 88.88 & 88.88 &  3.73 &  2.82 &  2.43 &  1.52 &  6.48 &  5.32 &  3.57 &  1.93 &   1.16 &   3.39 &   1.74 & 111 &   \\
                            11            &  17 39 37.28& $-30  $  08 51.64  &   10.02              &     7.90              &     6.67              & 16.94 &  9.99 &  6.63 &  4.57 &  3.09 & 88.88 & 88.88 & 88.88 & 88.88 &  4.24 &  3.14 &  3.02 & 98.98 &  4.99 &  4.55 &  3.34 &  2.14 &   2.06 &  $..$ &  $..$ & 111 &   \\
                            12            &  17 40 57.23& $-29  $  45 31.45  &   11.87              &     8.87              &  $( 7.48           )$ & $..$ & 11.60 &  7.16 &  4.07 &  2.70 &  6.71 &  6.18 &  5.14 &  4.46 &  4.30 &  3.20 &  2.80 & 98.98 &  7.80 &  6.50 &  3.97 &  2.27 &   1.94 &   4.08 &   2.79 & 111 &   \\
                            13            &  17 41 16.81& $-31  $  38 10.59  &   14.04              &    10.49              &     8.43              & $..$ & 13.54 &  8.30 &  4.98 &  3.05 &  6.79 &  6.00 &  5.48 &  4.79 &  3.97 &  2.78 &  2.53 & 98.98 &  6.97 &  5.42 &  3.32 &  1.70 &   1.50 &  96.96 &   2.61 & 111 &   \\
                            14            &  17 41 26.93& $-29  $  30 46.97  &   12.00              &     9.17              &     7.39              & $..$ & 11.80 &  7.39 &  4.48 &  3.56 & 88.88 & 88.88 & 88.88 & 88.88 &  5.01 & 98.98 & 98.98 & 98.98 &  5.87 &  5.07 &  3.59 &  2.18 &   2.86 &  $..$ &  $..$ & 191 &   \\
                            15            &  17 41 31.31& $-30  $  00 18.87  &    7.73              &     5.73              &     4.64              & 16.43 &  8.61 &  6.36 &  3.24 &  2.03 & 88.88 & 88.88 & 88.88 & 88.88 &  3.13 &  1.97 &  1.88 & 98.98 &  4.34 &  3.73 &  2.71 &  1.92 &   1.59 &  $..$ &  $..$ & 111 &   \\
                            16            &  17 41 36.86& $-29  $  29 30.99  &   12.69              &     9.53              &     7.58              & $..$ & 12.23 &  7.06 &  3.79 &  2.00 &  5.42 & $..$ &  4.59 &  4.28 &  3.48 &  2.41 &  2.15 &  1.45 &  5.86 &  4.90 &  2.90 &  1.47 &   1.93 &  $..$ &  $..$ & 111 &   \\
                            17            &  17 41 37.41& $-29  $  32 05.73  &   12.18              &     9.08              &     7.30              & $..$ & 12.26 &  7.37 &  4.60 &  2.85 & 88.88 & 88.88 & 88.88 & 88.88 &  4.49 &  3.44 &  3.07 & 98.98 &  5.44 &  4.50 &  3.41 &  2.08 &   2.51 &  $..$ &  $..$ & 111 &   \\
                            18            &  17 42 04.36& $-29  $  58 46.37  &   13.12              &     9.88              &     8.06              & $..$ & 12.16 &  7.33 &  4.83 &  3.40 &  6.68 &  6.35 &  5.59 &  5.31 &  5.26 & 98.98 & 98.98 & 98.98 &  6.47 &  5.88 &  4.79 &  3.64 &   3.40 &   4.97 &  96.96 & 111 &   \\
                            19            &  17 42 06.86& $-28  $  18 32.39  &    9.64              &     7.89              &     6.88              & 15.86 &  9.59 &  6.87 &  4.82 &  3.12 & 88.88 & 88.88 & 88.88 & 88.88 &  4.66 &  3.57 &  3.56 &  3.20 &  6.14 &  5.58 &  3.90 &  2.55 &   2.75 &   4.70 &   3.54 & 111 &   \\
                            20            &  17 42 23.28& $-29  $  39 35.57  &   12.13              &     9.13              &     7.46              & $..$ & 12.52 &  7.60 &  5.11 &  2.95 & 88.88 & 88.88 & 88.88 & 88.88 &  4.43 &  3.29 &  3.03 & 98.98 &  5.85 &  5.01 &  3.70 &  2.27 &   2.19 &   4.75 &   2.83 & 111 &   \\
                            21${\rm ^*}$  &  17 42 32.91& $-29  $  41 25.10  &   11.62              &     9.15              &     7.76              & 17.23 & 11.78 &  7.79 & $..$ &  4.64 &  6.94 &  6.47 &  6.08 &  5.77 & 97.97 & 97.97 & 97.97 & 97.97 &  6.51 &  6.08 &  5.46 &  3.80 &   3.70 &  $..$ &  $..$ & 121 &   \\
                            22            &  17 42 32.48& $-29  $  41 10.73  &   12.55              &     9.46              &     7.67              & $..$ & 11.94 &  7.19 &  4.33 &  3.08 & 88.88 & 88.88 & 88.88 & 88.88 & 97.97 & 97.97 & 97.97 & 98.98 &  6.54 &  4.72 &  3.80 &  2.24 &   2.23 &  96.96 &   2.61 & 111 &   \\
                            23            &  17 42 44.87& $-30  $  04 08.08  &    9.10              &     6.87              &     5.67              & 16.27 &  8.65 &  6.40 &  4.29 &  2.71 &  7.21 & $..$ &  4.42 &  4.20 &  4.30 &  3.36 &  3.16 & 98.98 &  4.98 &  4.52 &  3.59 &  2.82 &   2.98 &  $..$ &  $..$ & 111 &   \\
                            24            &  17 43 09.81& $-29  $  24 03.32  &   15.54              &    10.79              &     8.19              & $..$ & 14.50 &  7.53 &  3.65 &  1.93 &  5.29 & $..$ &  4.11 &  3.71 &  3.40 &  2.12 &  1.72 &  1.01 &  5.03 &  3.72 &  2.29 &  0.79 &   1.14 &  96.96 &   1.55 & 111 &   \\
                            25            &  17 43 23.46& $-28  $  53 50.33  &    8.44              &     6.21              &     4.97              & 15.73 &  8.01 & 93.93 &  2.87 &  2.16 &  4.05 &  3.85 &  3.59 &  3.20 &  3.30 &  2.41 &  2.13 &  2.17 &  3.73 &  3.13 &  2.50 &  1.78 &   2.10 &  96.96 &   2.08 & 111 &   \\
                            26            &  17 43 25.26& $-29  $  45 28.56  &   15.88              &    11.69              &     9.09              & $..$ & 15.23 &  8.99 &  4.92 &  2.76 &  6.80 &  6.14 &  5.12 &  4.55 &  4.16 &  2.94 &  2.58 & 98.98 & 94.94 & 94.94 &  3.15 &  1.62 &   1.43 &  96.96 &   2.53 & 111 &   \\
                            27            &  17 43 32.72& $-29  $  15 39.37  &   13.34              &     9.63              &     7.47              & $..$ & 12.06 &  6.91 &  3.94 &  2.45 &  5.37 &  5.21 &  4.50 &  4.18 &  3.50 &  2.50 &  2.33 &  1.85 & 94.94 & 94.94 &  2.91 &  1.45 &   1.85 &  96.96 &   2.41 & 111 &   \\
                            28            &  17 43 33.13& $-29  $  51 33.11  &   15.02              &    10.83              &     8.42              & $..$ & 15.00 &  8.41 &  4.68 &  2.88 &  6.89 &  6.34 &  4.78 &  4.34 &  4.70 &  3.60 &  3.30 & 98.98 &  6.38 &  5.34 &  4.33 &  2.92 &   2.62 &  96.96 &   2.96 & 111 &   \\
                            29            &  17 43 34.79& $-29  $  40 30.41  & $(16.90           )$ &    13.15              &     9.76              & $..$ & $..$ &  9.15 &  4.65 &  2.96 & 88.88 & 88.88 & 88.88 & 88.88 &  4.65 &  3.42 &  2.78 & 98.98 &  5.37 &  4.49 &  3.98 &  2.30 &   1.62 &  96.96 &   2.90 & 111 &   \\
                            30            &  17 43 35.12& $-29  $  24 47.22  & $(18.47           )$ &    12.22              &     9.09              & $..$ & $..$ &  8.82 &  4.97 &  2.95 & 88.88 & 88.88 & 88.88 & 88.88 &  4.57 &  3.13 &  3.08 & 98.98 &  7.61 &  6.05 &  4.41 &  2.63 &   2.01 &  $..$ &  $..$ & 111 &   \\
\hline
\end{tabular}
\begin{list}{}{}
\item[{\bf Notes.}]  Identification numbers are from Tables \ref{table:detections} and \ref{table:nondetections} of this work,
and from Tables 2 and 3 of \citet{messineo02} (see also Table 2 in \citet{messineo04}).
Magic values are set as follow: $..$ = general indicator of a missing entry; 
98.98 = upper limits;  97.97 = blending ( (i) automatically identified by having found multiple WISE stars
or multiple GLIMPSE bright stars at this position; (ii) datapoints removed after visual image inspection
because of astrometric offsets due to stellar blending or strong nebular emission); 
96.96 = area not covered by the survey; 95.95 = faint source not detected; 
94.94 = datapoints removed because very offset from the SED defined by the other datapoints; 
93.93 = magnitude above the saturation limit;
88.88 = no magnitude was found, but the corresponding image is  saturated (bright target).
For 2MASS $JH$\Ks\ bands, upper limits are given between parenthesis.\\
($\dag$) 2MASS $JH$\Ks\ magnitudes are missing or blended in 2MASS. UKIDSS coordinates and magnitudes are provided, 
between parenthesis when affected by saturation.\\
($*$) These marked Id indicate targets for which multiple WISE sources fall within the beam. 
There is a small chance that the selected counterparts are not the actual masing stars.\\
($+$) Nstar is a flag that concatenates the number of MSX point sources detected within a radius  of 14\farcs5 (first digit), the
number of WISE point sources detected within a radius of 10\farcs0 (second digit), 
and the number of preselected GLIMPSE point sources within a radius of 10\farcs0 (third digit). 
For example, Nstar=111 means that in the searched circle we found only 1 MSX star, 1 WISE, and 1 bright GLIMPSE.
Exception: when the second digit is 9, it means that no WISE target was found in the AllWISE Data Release, and the adopted 
magnitudes come from the older  WISE All-Sky Data Release.~\\
$(\rm a)$ For star \#294, there are additional $JHK$ measurements from \citet{cabrera06} 
that well agree with the 2MASS ($J=11.333, H=7.346, K=5.442$ mag). This suggests that 
the DENIS \Ks\ measurement (7.73 mag) could be affected by artifact.~\\
$(**)$ The full 
table contains 571 objects and is available in electronic form at the CDS via anonymous ftp to
cdsarc.u-strasbg.fr (130.79.128.5) or via http://cdsweb.u-strasbg.fr/cgi-bin/qcat?J/A+A/xxx/xxx.
\end{list}
}
\end{center}
\end{sidewaystable*}

\end{document}